\documentclass{article}

\usepackage{amsthm}
\usepackage{amsmath}
\usepackage{amssymb}
\usepackage{dlfltxbcodetips}
\usepackage{gastex}
\usepackage{subcaption} 
\usepackage{color}
\usepackage{booktabs}
\usepackage{multirow}
\usepackage{hyperref}
\usepackage{arxiv}
\usepackage{url}
\usepackage{natbib}

\theoremstyle{definition}

\begin{document}

\numberwithin{equation}{section}
\numberwithin{theorem}{section}

\title{Simultaneous confidence intervals for an extended Koch-R\"ohmel design in three-arm non-inferiority trials}
\author{Martin Scharpenberg\thanks{Correspondence to: Martin Scharpenberg,\ \ \href{mailto:mscharpenberg@uni-bremen.de}{mscharpenberg@uni-bremen.de},\ \ \url{https://orcid.org/0000-0003-1584-0056}}\\
  Competence Center for\\
	Clinical Trials Bremen\\
  University of Bremen\\
  Bremen, Germany \\  
   \And
  Werner~Brannath \\
  Institute for Statistics and\\ 
  Competence Center for\\
  Clinical Trials Bremen\\
  University of Bremen\\
  Bremen, Germany \\}
\date{\today}

\maketitle

\begin{abstract}
Three-arm `gold-standard' non-inferiority trials are recommended for indications where only unstable reference treatments are available and the use of a placebo group can be justified  ethically. For such trials several study designs have been suggested that use the placebo group for testing 'assay sensitivity', i.e.\ the ability of the trial to replicate efficacy. Should the reference fail
in the given trial, then non-inferiority could also be shown with an ineffective experimental treatment and hence becomes useless. In this paper we extend the so called Koch-R\"ohmel design where a proof of efficacy for the experimental treatment is required in order to qualify the non-inferiority test. While efficacy of the experimental treatment is an indication for assay sensitivity, it does not guarantee that the reference is sufficient efficient to let the non-inferiority claim be meaningful. It has therefore been suggested to adaptively test non-inferiority only if the reference demonstrates superiority to placebo and otherwise to 
test $\delta$-superiority of the experimental treatment over placebo, where $\delta$ is chosen in such a way that it provides proof of non-inferiority  with regard to the reference's historical effect. In this paper we extend the previous work by complementing its adaptive test with compatible simultaneous confidence intervals. 
Confidence intervals are commonly used and suggested by regulatory guidelines for non-inferiority trials.
We show how to adopt different approaches to simultaneous confidence intervals from the literature to the setting of three-arm non-inferiority trials and compare these methods in a simulation study. Finally we apply these methods to a real clinical trial example.
\end{abstract}

\section{Introduction}

We consider three-arm trials including an experimental treatment ($E$), an established reference ($R$) and a~placebo (see, e.g., \citealp{KR05}). Typical indications for such trials are for example asthma or depression, where the use of placebo is acceptable from an ethical point of view and where the effect of the reference has normally a~high variance. New treatments are desired to obtain better effects for certain patient collectives or to reduce side effects. An effect that is not inferior to that of the reference is then regarded as successful. To assure assay sensitivity, a~placebo group is included in the trial.

Denote by $\mu_i$ the effect in group~$i$, where $i\in\{E,R,P\}$. We will use the following notations for the hypotheses of interest:
\begin{align*}
H_{EP}^S &: \mu_E - \mu_P \le 0 \quad \text{(to show superiority of $E$ over $P$)} \\
H_{ER}^N &: \mu_E - \mu_R \le -\delta_0 \quad \text{(to show non-inferiority of $E$ compared to $R$ with margin $\delta_0 > 0$)} \\
H_{EP}^{\delta_1} &: \mu_E - \mu_P \le \delta_1 \quad \text{(to show superiority by $\delta_1 > 0$ of $E$ over $P$)} 
\end{align*}
Without loss of generality, we can assume that $\mu_P=0$. The non-inferiority margin is normally chosen as $\delta_0 = r\mu_R^h$, where $r\in(0,1)$ and $\mu_R^h$ is the historical reference effect, i.e., the effect observed in earlier studies. A~common choice is $r=1/2$ or smaller. We will come to the choice of~$\delta_1$ immediately.

The gold-standard for testing in three-arm trials was proposed by \citealp{KR05} and consists of a~hierarchical procedure rejecting first $H_{EP}^S$ to assure assay sensitivity and then rejecting $H_{ER}^N$ to confirm efficacy of the new treatment. There are several possibilities to proceed after the second rejection. An important issue is the uncertainty about the strength of the reference in the study. If actually $\mu_R=0$, then the rejection of $H_{ER}^N$ is both easy and worthless (see \citealp{HP05}). Since it is unsatisfying to spend level on showing superiority of the reference over placebo, an idea is to consider the hypothesis $H_{EP}^{\delta_1}$ instead. This targets the clinical relevance of the new treatment and implicitly assures non-inferiority to the historical reference effect. We propose to choose $\delta_1= (1-r)\mu_R^h$. If the effect of the reference in the study corresponds to the historical effect, i.e., $\mu_R=\mu_R^h$, then non-inferiority implies $\mu_E > \mu_R - \delta_0 = (1-r)\mu_R^h$. On the other hand, if the reference in the study is weak, then the rejection of $H_{EP}^{\delta_1}$ assures likewise that $\mu_E > \delta_1= (1-r)\mu_R^h$. Hence, non-inferiority to the (historical) reference with margin $r\mu_R^h = \delta_0$ is shown. 

In \citealp{BS15}, a~flexible extension of the Koch-R\"ohmel design was proposed, which is in the spirit of the above argumentation. A~hierarchical test of the sequence $H_{EP}^S$, $H_{ER}^N$, $H_{EP}^{\delta_1}$ was introduced, and success of the study was interpreted in dependence of the reference strength. If ``$R>P$'', then success is reached if $H_{ER}^N$ is rejected. If ``$R\le P$'', then success means rejecting~$H_{EP}^{\delta_1}$. The test of the hypothesis $H_{RP}^S: \mu_R - \mu_P \le 0$ served as a~filter (``$R>P$'' or ``$R\le P$'') for the interpretation of the study results. It was shown that the overall probability for erroneously declaring success can be controlled at level $\alpha$ without a~confirmatory test of the reference effect.

It is well known that confidence intervals provide more information than hypothesis testing. They are particularly important for non-inferiority trials. For example, we can quantify the amount by which $\mu_E$ is superior to $\mu_P$, or even superior to $\mu_P+\delta_1$. We may also be able to improve the non-inferiority to the reference or even show superiority, i.e., prove that $\mu_E - \mu_R > 0$. In this paper, we extend the idea of \citealp{BS15} by introducing three sets of simultaneous confidence intervals (SCIs) for the effects $\mu_E-\mu_P$ and $\mu_E-\mu_R$. The first set of proposed SCIs is based on stepwise confidence intervals introduced in \citealp{HB99}. We will find an intrinsic filter describing the strength of the reference, which is perfectly compatible with the interpretation of these simultaneous confidence bounds.  Thus, we obtain an intuitive procedure to decide about the success of the study and obtain information via a~SCI. Another set of SCIs will be based on the work of \citealp{SB14}, who introduced informative confidence intervals in hierarchical testing. Finally, we will compare these two sets of SCIs to the single step SCIs which can easily be derived for the aforementioned effects.

The paper is organized as follows. We introduce the simultaneous confidence intervals in Section~\ref{sec_SCIS} and give numerical results on their performance in Section~\ref{sec_numerical}. We apply the different intervals to data of clinical trial in Section~\ref{sec_example} and end with a~discussion in Section~\ref{sec_discussion}.

\section{Description of the SCIs}\label{sec_SCIS}

\subsection{SCIs with IU filter} \label{sec_design}

We assume that all observations are normally distributed with common standard deviation $\sigma>0$. Denote by $n_i$ the sample size of group $i\in\{E,R,P\}$. Then the observed mean in group~$i$ is $X_i\sim N(\mu_i,\sigma^2/n_i)$. Define the univariate confidence bounds
$$
\ell_{EP} = X_E - X_P -z_\alpha \sigma \sqrt{n_E^{-1}+n_P^{-1}} \quad \text{and} \quad
\ell_{ER} = X_E - X_R -z_\alpha \sigma \sqrt{n_E^{-1}+n_R^{-1}},
$$
where $z_\alpha=\Phi^{-1}(1-\alpha)$ is the quantile of the standard normal distribution. According to \citealp{HB99}, simultaneous lower confidence bounds for $\theta=(\theta_1,\theta_2)=(\mu_E-\mu_P,\mu_E-\mu_R)$ with coverage probability $(1-\alpha)$ are given by
\begin{equation}
\label{lep}
L_{EP} =\begin{cases}
	  \ell_{EP} & \text{if } \ell_{EP} < 0, \\
	  0     & \text{if } \ell_{EP} \ge 0,\ \ell_{ER} < -\delta_0, \\
	  L_{\min}:=\min\{\ell_{EP},\ell_{ER}+\delta_0\} & \text{if } \ell_{EP} \ge 0,\ \ell_{ER} \ge -\delta_0,
      \end{cases}
\end{equation}
and
\begin{equation}
\label{ler}
L_{ER} =\begin{cases}
	  -\infty & \text{if } \ell_{EP} < 0, \\
	  \ell_{ER}  & \text{if } \ell_{EP} \ge 0,\ \ell_{ER} <  -\delta_0, \\
	  L_{\min}-\delta_0=\min\{\ell_{EP}-\delta_0,\ell_{ER}\} & \text{if } \ell_{EP} \ge 0,\ \ell_{ER} \ge -\delta_0.
      \end{cases}
\end{equation}
These SCIs are compatible with the rejection decisions of the Koch-R\"ohmel design, i.e., with a~hierarchical test of $H_{EP}^S$ and~$H_{ER}^N$ as in \citealp{MHL95}. The first line of \eqref{lep} and~\eqref{ler} corresponds to the case where the hypothesis $H_{EP}^S$ cannot be rejected. In this case the study is considered a failure. The second line is the case where $H_{EP}^S$ is rejected, but $H_{ER}^N$ is not. We also consider this case as failure, even though non-inferiority is not of interest in the case of a~weak reference. However, if the reference fails then $H_{ER}^N$ should be rejected easily. The most important situation occurs in the third line of \eqref{lep} and~\eqref{ler}, where both hypotheses $H_{EP}^S$ and $H_{ER}^N$ are rejected. The confidence bounds $L_{EP}$ and $L_{ER}$ are then both determined by either information on the effect $\mu_E-\mu_P$ via $\ell_{EP}$ or on the effect $\mu_E-\mu_R$ via $\ell_{ER}$, where the boundary between these two options can easily be calculated. Under the last condition of \eqref{lep} and~\eqref{ler}:
\begin{equation}
\label{filterE}
L_{\min}=\ell_{ER} + \delta_0 \iff X_R - X_P \ge z_\alpha\sigma\big(\sqrt{n_E^{-1}+n_P^{-1}} - \sqrt{n_E^{-1}+n_R^{-1}}\big) + \delta_0.
\end{equation}
It appears unnatural at first sight that $L_{ER}$, which is a~lower bound for~$\theta_1$, is given by an estimate of~$\theta_2$ in some cases, and vice versa. However, there is a~transparent way of interpreting the bounds: If the reference is ``strong'' in the sense of~\eqref{filterE}, then the confidence bound for the effect of $E$ versus~$R$ is of interest, which is reflected by the fact that $L_{\min}=\ell_{ER} + \delta_0$ and hence the level is fully exploited for the parameter $\mu_E-\mu_R$. In the other case, where the reference is ``weak'' so that the inequality in~\eqref{filterE} is not satisfied, then $L_{\min}=\ell_{EP}$, i.e., the level is exploited for $\mu_E-\mu_P$. This makes sense, because the difference of the new treatment to the weak reference is not of interest and non-inferiority to the historical reference is targeted indirectly via a~bound for the effect of $E$ versus~$P$. In summary, the study can be judged as successful if either the reference is strong and $L_{ER} \ge -\delta_0=-r\mu_R^h$ or the reference is weak and $L_{EP} \ge \delta_1=(1-r)\mu_R^h$. This inference can be made without needing to test the reference effect directly. Thus, Equation~\eqref{filterE} serves as an intrinsic filter for the interpretation of the study results.

Formally, the definition of $L_{\min}$ is an application of nested intersection union tests of hypotheses $H_\Lambda^\vartheta: \Lambda \le \vartheta$ at level~$\alpha$ for the parameter $\Lambda=\min\{\mu_E-\mu_P,\mu_E-\mu_R+\delta_0\}$. The value of $\Lambda$ equals $\mu_E-\mu_P$ if and only if $\mu_R \le \mu_P+\delta_0$, otherwise it quantifies $\mu_E-\mu_R+\delta_0$. This corresponds to the philosophy of a~weak versus strong reference. Since the SCIs \eqref{lep} and~\eqref{ler} and the filter \eqref{filterE} are the application of nested intersection union tests, we will call the filter \textit{IU filter}.

\begin{figure}
\begin{center}
\includegraphics[width=0.8 \textwidth]{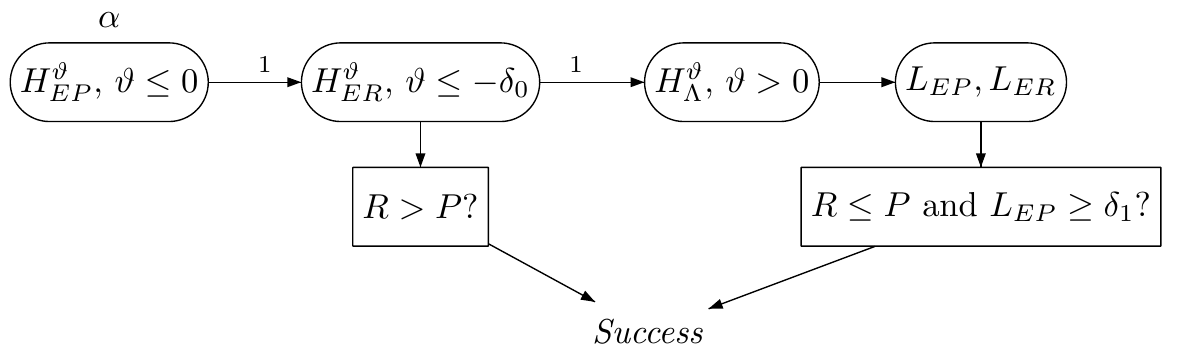}
\end{center}
\caption{\label{fig_design_formal} Formal description of the simultaneous confidence intervals. The hypotheses are $H_{EP}^{\vartheta}:\mu_E-\mu_P \le\vartheta$, $H_{ER}^{\vartheta}:\mu_E-\mu_R \le\vartheta$ and $H^\vartheta_\Lambda:\Lambda\le\vartheta$ for $\Lambda=\min\{\mu_E-\mu_P,\mu_E-\mu_R+\delta_0\}$. The IU filter ``$R>P$'' is given in~\eqref{filterE}.}
\end{figure}

\begin{figure}
\begin{center}
\includegraphics[width=0.9 \textwidth]{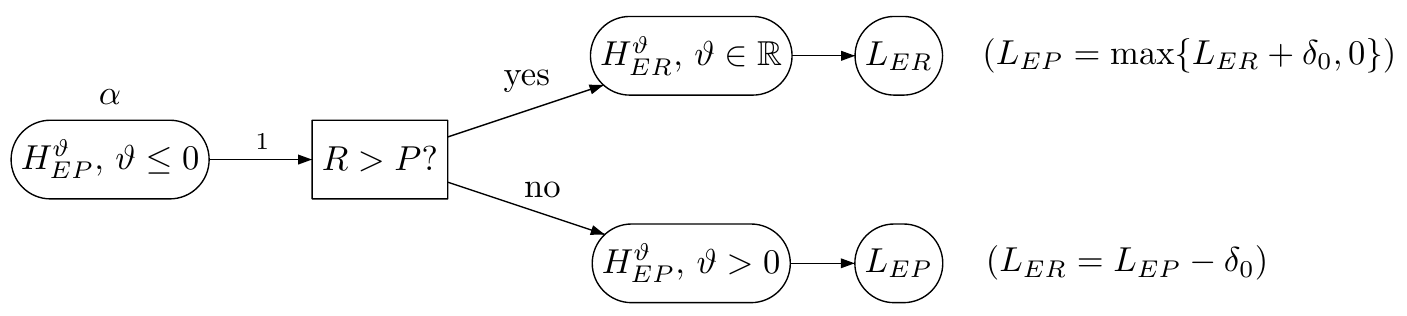}
\end{center}
\caption{\label{fig_design_intuitive} Intuitive graphical description and interpretation of the simultaneous confidence intervals. Notations as in Figure~\ref{fig_design_formal}.}
\end{figure}

The procedure of defining simultaneous confidence intervals and applying them to define the success of the study can be illustrated graphically using the notation of \citealp{SB14}. This is done in Figure~\ref{fig_design_formal}. The test of the hypothesis family $(H_{EP}^\vartheta)_{\vartheta\le 0}$ (at level~$\alpha$) leads to a~confidence bound $L_{EP}=\ell_{EP} < 0$ if $H_{EP}^S$ cannot be rejected, or $L_{EP} = 0$ otherwise. In the first case, the procedure stops with $L_{ER}=-\infty$, while in the latter case, the family of hypotheses $(H_{ER}^\vartheta)_{\vartheta\le -\delta_0}$ is tested (at level~$\alpha$), resulting in a~confidence bound $-\infty < L_{ER} \le -\delta_0$. If $L_{ER}=-\delta_0$, i.e., if $H_{ER}^N$ is rejected, then the level~$\alpha$ is shifted further to the hypothesis family $(H_{\Lambda}^\vartheta)_{\vartheta > 0}$, improving the confidence bounds~$L_{EP}$ and $L_{ER}$ simultaneously.

This procedure modifies the hierarchical test design introduced in \citealp{BS15}, where the hypothesis $H_{EP}^{\delta_1}$ is tested after rejection of $H_{EP}^S$ and $H_{ER}^N$. Here, more information about the size of both effects $\mu_E-\mu_P$ and $\mu_E-\mu_R$ can be gained via a~confidence interval for~$\Lambda$. Another difference is that the filter ``$R<P$'' is now given via \eqref{filterE}, while it was a level-$\alpha$-test of the hypothesis $H_{RP}:\mu_R-\mu_P \le 0$ in \citealp{BS15}. The new filter has a remarkable advantage. Indeed, it was discussed in \citealp{BS15} that the rather complex strategy in Figure~\ref{fig_design_formal} can be replaced by an intuitive strategy for many situations. Actually, we can give an intuitive picture for \textit{all} situations if we use the new filter. The corresponding graph is shown in Figure~\ref{fig_design_intuitive}. After rejection of the gatekeeper $H_{EP}^S$, it is decided via the filter whether to look at the lower bound for $\mu_E-\mu_R$ or the lower bound for $\mu_E-\mu_P$. In the first case $R>P$, the study is successful if $L_{ER}:=\ell_{ER}\ge -\delta_0$, in the second case, $R<P$, it is successful if $L_{EP}:=\ell_{EP}\ge \delta_1$. Additionally, one can report $L_{EP}=\max\{0,L_{ER}+\delta_0\}$ in the first case, or respectively, $L_{ER}=L_{EP}-\delta_0$ in the second case. Although the procedure in Figure~\ref{fig_design_intuitive} is data-driven via the filter, one can show that it maintains the given coverage probability~$1-\alpha$, because it is equivalent to the procedure in Figure~\ref{fig_design_formal}, which is based on the intervals from \citealp{HB99} (see Appendix). Therefore, also the probability of erroneously stating success of the study is controlled.

\subsection{Informative SCIs}\label{sec_informative}
Next, we present another way to calculate simultaneous lower confidence bounds for $(\mu_E-\mu_P,\mu_E-\mu_R)$ with coverage probability $(1-\alpha)$ by employing the method of \citealp{SB14}. They derived so called informative SCIs in hierarchical testing. A confidence bound is informative, if it does not stick to the border of the null-hypothesis interval with positive probability, in case the null hypothesis is rejected.

\begin{figure}
\begin{center}
\includegraphics[width=0.8 \textwidth]{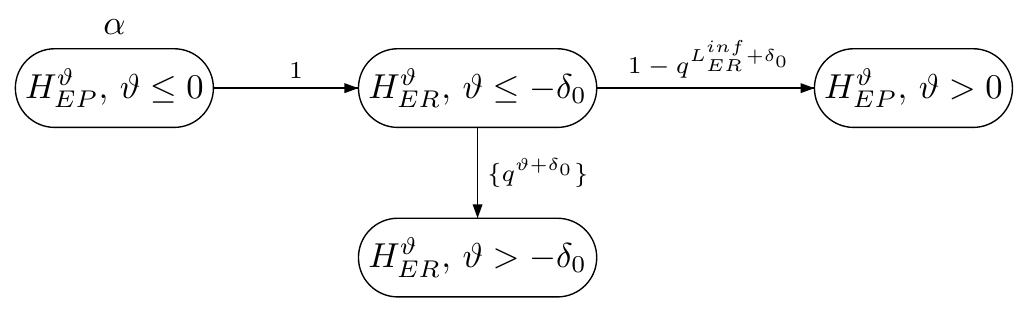}
\end{center}
\caption{\label{fig_informative_SCI} Graphical description of the algorithm to determine informative SCIs for $\mu_E-\mu_P$ and $\mu_E-\mu_R$, based on the notation and procedure of \citealp{SB14}.}
\end{figure}

The procedure for deriving the simultaneous lower confidence bounds, denoted by $L_{EP}^{inf}$ and $L_{ER}^{inf}$ in the following, is illustrated in Figure~\ref{fig_informative_SCI} and works as follows: First, fix a value $0<q<1$. Then, $H_{EP}^S:\mu_E-\mu_P\leq 0$ is tested at full level $\alpha$. If $H_{EP}^S$ cannot be rejected, the procedure stops and reports $L_{EP}^{inf}=\ell_{EP}$ and $L_{ER}^{inf}=-\infty$. If $H_{EP}^S$ can be rejected, the full level $\alpha$ is passed to testing $H_{ER}^N$. If $H_{ER}^N$ cannot be rejected, the procedure stops and reports $L_{EP}^{inf}=0$ and $L_{ER}^{inf}=\ell_{ER}$. If however, also $H_{ER}^N$ is rejected, we test each $H_{ER}^\vartheta:\mu_E-\mu_R\leq\vartheta$ at level $q^{\vartheta+\delta_0}\alpha$. The lower confidence bound $L_{ER}^{inf}$ for $\mu_E-\mu_R$ will then be the first $\vartheta$ for which $H_{ER}^\vartheta$ cannot be rejected at the respective level. It can be shown that this bound is the unique solution to 

\begin{equation*}
1-\Phi\left\{\frac{X_E-X_R-\vartheta}{\sigma\sqrt{n_E^{-1}+n_R^{-1}}}\right\}=q^{\vartheta+\delta_0}\alpha.
\end{equation*}

In the next step, we pass the remaining $\alpha$-level ($1-q^{L_{ER}^{inf}+\delta_0}\alpha$) to the hypotheses $H_{EP}^\vartheta$ and determine the lower confidence bound for $\mu_E-\mu_P$ as $L_{EP}^{inf}=\max\left\{0,\tilde\ell_{EP}\right\}$, where $$\tilde\ell_{EP}= X_E - X_P -z_{q^{L_{ER}^{inf}+\delta_0}\alpha} \sigma \sqrt{n_E^{-1}+n_P^{-1}}.$$ Note, that in the cases where $H_{EP}^S$ or $H_{ER}^N$ cannot be rejected, these SCIs coincide with those presented in the previous section. Only in the case, where E is superior to P and non-inferior to R, the two procedures to calculate SCIs differ.

Note, that the intervals introduced here are not informative in the sense indicated at the beginning of this subsection. This is due to the fact, that if $H^S_{EP}$ is rejected, but $H_{ER}^N$ is not, which might be the case with positive probability, the confidence bound for $\mu_E-\mu_P$ is 0 by construction. However, since the SCIs introduced here are based on the informative SCIs introduced by \citealp{SB14}, we will refer to them as \textit{informative} SCIs in the following.

While the SCIs introduced in section~\ref{sec_design} offered an intrinsic filter to determine whether ``$R>P$'' or ``$R\le P$'', this is not the case for the SCIs introduced in this section. For the interpretation of the confidence intervals respectively the study results, we will use the \textit{superiority filter} presented in \citealp{BS15}, which declares ``$R>P$'' if 
\begin{equation}
\frac{X_R-X_P}{\sigma\sqrt{n_R^{-1}+n_P^{-1}}}\geq z_\alpha,
\label{supfilter}
\end{equation}
i.e.~if $H_{RP}^S$ can be rejected. The interpretation then works similar to the case of section~\ref{sec_design}. After rejection of $H_{ER}^S$ we can declare a successful study, if either ``$R>P$'' (i.e.~$H_{RP}^S$ can be rejected) and $L_{ER}^{inf}>-\delta_0$, or if ``$R\leq P$'' (i.e.~$H_{RP}^S$ cannot be rejected) and $L_{EP}^{inf}>\delta_1$. Furthermore, $L_{EP}^{inf}$ respectively $L_{ER}^{inf}$ can additionally be reported.

\subsection{Single-Step SCIs}\label{sec_singlestep}
We now briefly introduce simultaneous single step confidence intervals for the effects considered. To this end we consider the test statistics 
\begin{align*}
T_{EP} = \frac{X_E-X_P}{\sigma\sqrt{n_E^{-1}+n_P^{-1}}}, \quad
T_{ER}^N =\frac{X_E-X_R + \delta_0}{\sigma\sqrt{n_E^{-1}+n_R^{-1}}}.
\end{align*}
For large sample sizes $n_E, n_R$ and $n_P$ and for $\mu_E=\mu_P$ and $\mu_E-\mu_R=-\delta_0$ (i.e.~$H_{EP}^S\cap H_{ER}^N$ is true), the vector of test statistics $(T_{EP},T_{ER}^N)^\prime$ is asymptotically multivariate normally distributed with mean vector $\mu=(0,0)$ and covariance matrix $\Sigma=\begin{pmatrix}
	1 &  \rho\\
	\rho & 1
\end{pmatrix}$ with $\rho=\sqrt{\frac{c_Pc_R}{(1+c_P)(1+c_R)}}$ for $c_R=n_R/n_E$ and $c_P=n_P/n_E$. Let $d_\alpha$ be the equicoordinate $1-\alpha$ quantile of this distribution. With this definition, it is well known (cf.~\citealp{H96}) that $$
L_{EP}^S = X_E - X_P -d_\alpha \sigma \sqrt{n_E^{-1}+n_P^{-1}} \quad \text{and} \quad
L_{ER}^S= X_E - X_R -d_\alpha \sigma \sqrt{n_E^{-1}+n_R^{-1}},
$$ 
give simultaneous lower confidence bounds for $(\mu_E-\mu_P,\mu_E-\mu_R)$ with coverage probability $(1-\alpha)$. As for the intervals introduced in section~\ref{sec_informative} we need a filter for the interpretation of study results, when using the single-step SCIs introduced here. We will also use the superiority filter \eqref{supfilter} and declare success of the study if either ``$R>P$'' (i.e.~$H_{RP}^S$ can be rejected) and $L_{ER}^{S}>-\delta_0$, or if ``$R\leq P$'' (i.e.~$H_{RP}^S$ cannot be rejected) and $L_{EP}^{S}>\delta_1$.

\section{Properties of SCIs and probability of success} \label{sec_numerical}

\subsection{Optimal sample sizes}\label{sec_optimalN}
\citealp{BS15} presented a~method to calculate the minimal necessary sample size to obtain a~certain probability of success (i.e.\ probability of correctly declaring the study a ``success'')  which is based on a~procedure described in \citealp{SB13}. By ``success'' we mean that we can show either non-inferiority of the new treatment to the reference in the case where the filter is satisfied (denoted by ``Success~ER'' in the following), or to show $\delta_1$-superiority of the new treatment over placebo in the case where the filter is not satisfied (denoted by ``Success~EP''). The overall success probability (``Total PoS'') is the sum of Success~ER and Success~EP (for further explanation see section \ref{ex_interpretation}). The algorithm to determine the optimal sample sizes is based on the observation, that the hypotheses $H_{EP}^\vartheta:\mu_E-\mu_P\leq\vartheta, H_{ER}^\vartheta:\mu_E-\mu_R\leq\vartheta$ and $H_{RP}^\vartheta:\mu_R-\mu_P\leq\vartheta$ can be tested using test statistics which are multivariate normally distributed. Given specific sample size allocations (specified via the ratios $c_R=n_R/n_E$ and $c_P=n_P/n_E$) a probability of success, a significance level $\alpha$ and assumptions on the effects of E, R and P, the total sample size needed can then be derived via the cumulative distribution function of the respective multivariate normal distribution. In order to obtain the optimal sample sizes, the required total sample size N is optimized over the sample size allocations. 

The method presented in \citealp{BS15} can also be applied to our setup with the IU filter \eqref{filterE}. We want to compare the optimal sample sizes derived for this setup to those derived for the superiority filter \eqref{supfilter} To compare the two approaches, we make the following assumptions:
\begin{itemize}
\item desired power $90\%$ for the success of the study
\item one-sided FWER $\alpha=2.5\%$
\item independent normal observations with common standard deviation $\sigma=2$
\item non-inferiority margin $\delta_0=0.5$, which corresponds to half of the historical reference effect
\item the effect of the new treatment is equal to the historical reference effect, $\mu_{E} - \mu_P=1$
\end{itemize}

We consider three different scenarios for the effect of the reference over the placebo:

\begin{itemize}
\item Scenario 1: the reference is as good as observed historically, $\mu_{R} - \mu_P = 1$
\item Scenario 2: the reference is only half as good as observed historically, $\mu_{R} - \mu_P=0.5$
\item Scenario 3: the reference fails and is as good as placebo, $\mu_{R} - \mu_P=0$
\end{itemize}

Scenarios 2 and 3 are particularly relevant if the reference treatment is assumed to be ``unstable''. Unstable reference effects are observed in indications such as depression, where there is a rather large uncertainty about the mean treatment effect to be expected in a specific trial, inter alia due to individual variability in treatment effects (see also \citealp{EMA2010} and \citealp{Kaiser22}).

\begin{table}[h]
\centering
\caption{\label{tab_filter} Optimal sample sizes for the flexible non-inferiority design comparing the SCIs introduced in Section \ref{sec_SCIS} to the procedure of \citealp{BS15} without SCIs (see text for details)}
\begin{tabular}{cccccc}				
\toprule
	Scenario &&  Superiority filter & IU filter & Informative & Single Step \\
\midrule
 \multirow{4}{5pt}{1} & $n_E$ &	345	& 356 & 349 & 402 \\
											& $n_R$ &	350	& 348 & 348 & 406 \\
											& $n_P$ &	102 & 145 & 104 & 100 \\\cmidrule{2-6}
											& $N$ 	&	797  & 849 & 801 & 908 \\
\midrule
\multirow{4}{5pt}{2} 	& $n_E$ &	185	& 227 & 159 & 134 \\
											& $n_R$ &	182	& 75 	& 216 & 253 \\
											& $n_P$ &	303 & 285 & 313 & 323 \\\cmidrule{2-6}
											& $N$ 	&	670  & 587 & 688 & 710  \\
\midrule
\multirow{4}{5pt}{3} 	& $n_E$ &	341	& 306 & 348 & 397 \\
											& $n_R$ &	44	& 33 	& 52 	& 44 \\
											& $n_P$ &	339 & 325 & 346 & 399 \\\cmidrule{2-6}
											& $N$ 	&	724 & 661 & 746 & 840 \\
\bottomrule
\end{tabular}
\end{table}

\begin{figure}
\begin{center}
\includegraphics[width= .5\textwidth]{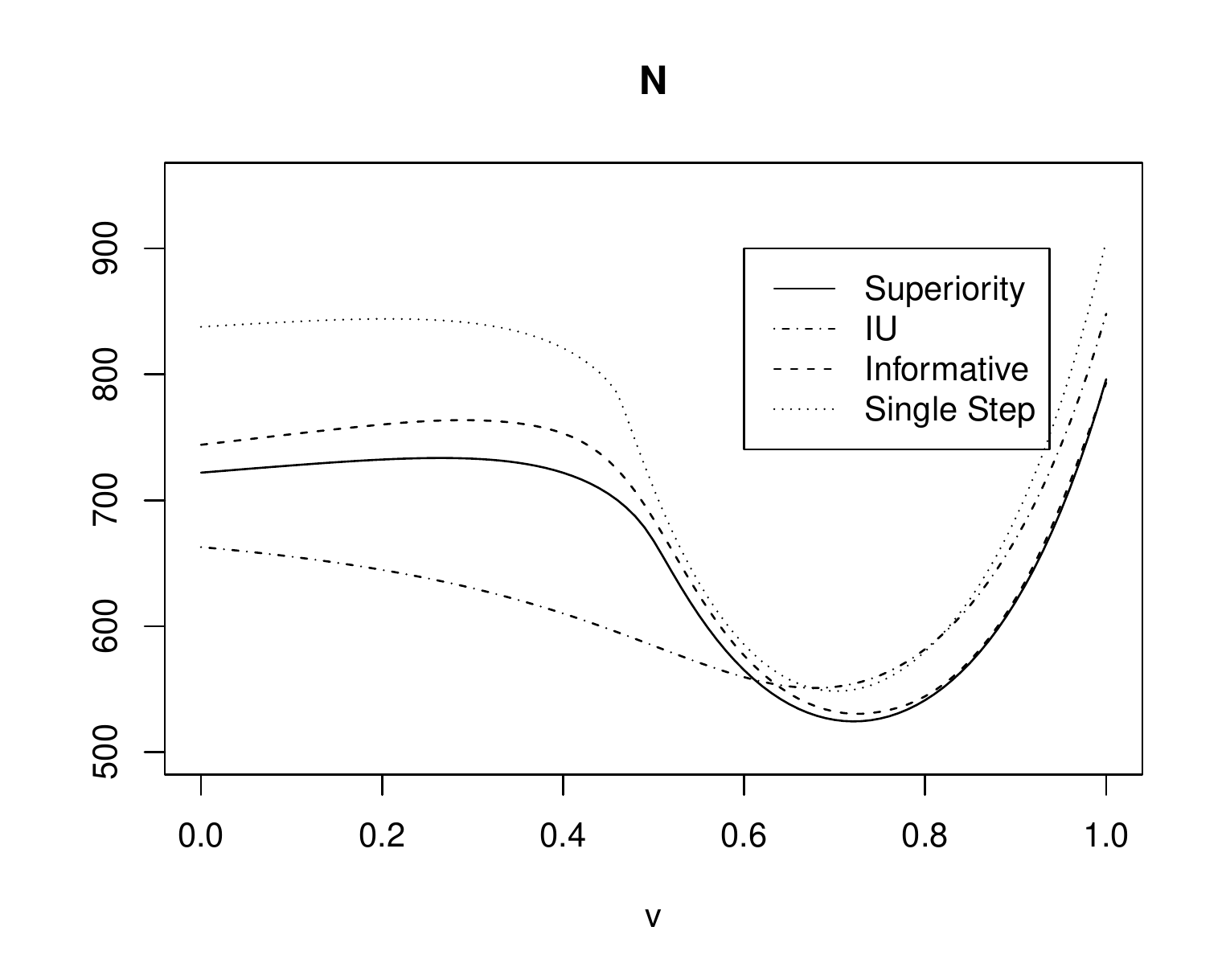}~\includegraphics[width= .5\textwidth]{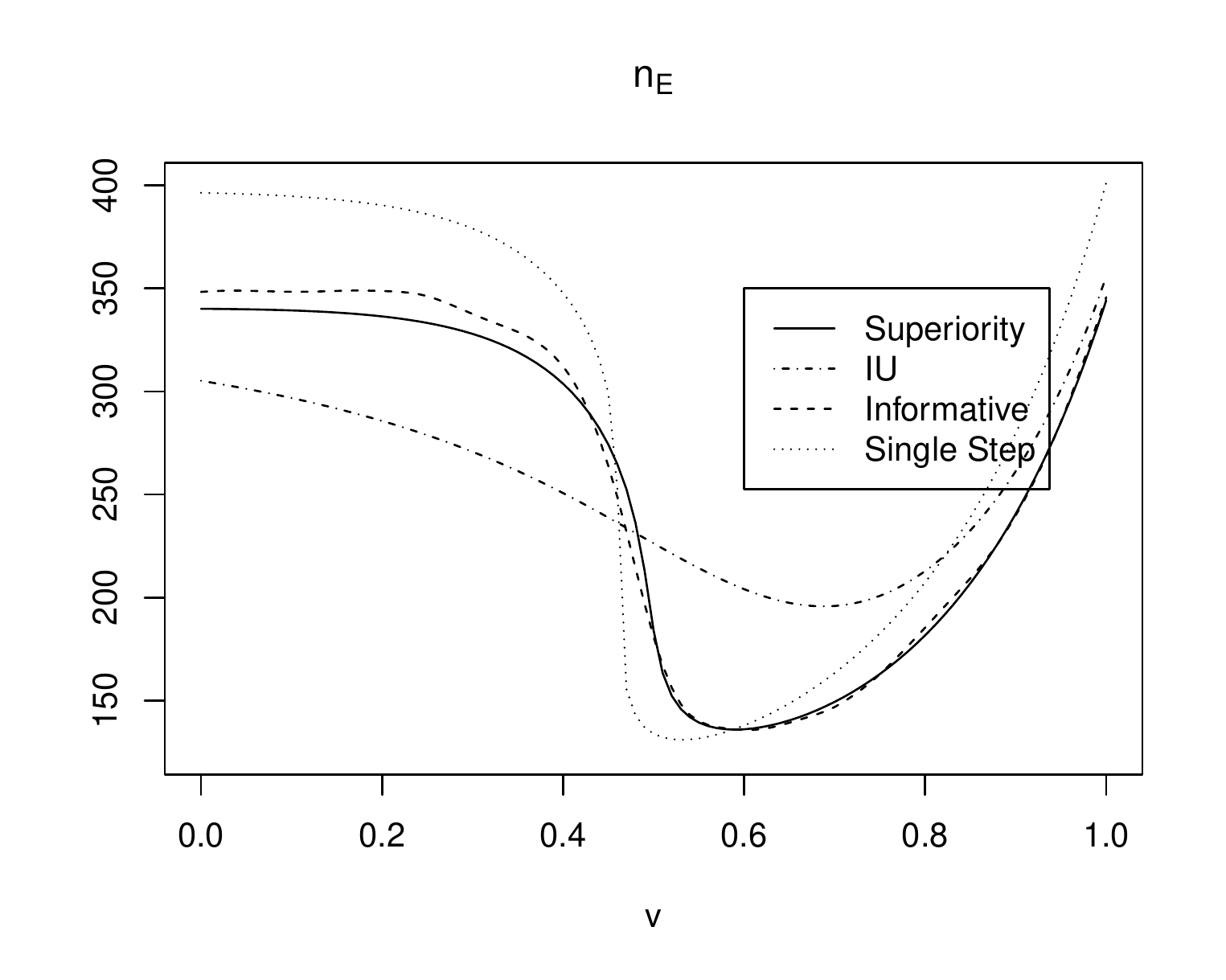}\\
\includegraphics[width= .5\textwidth]{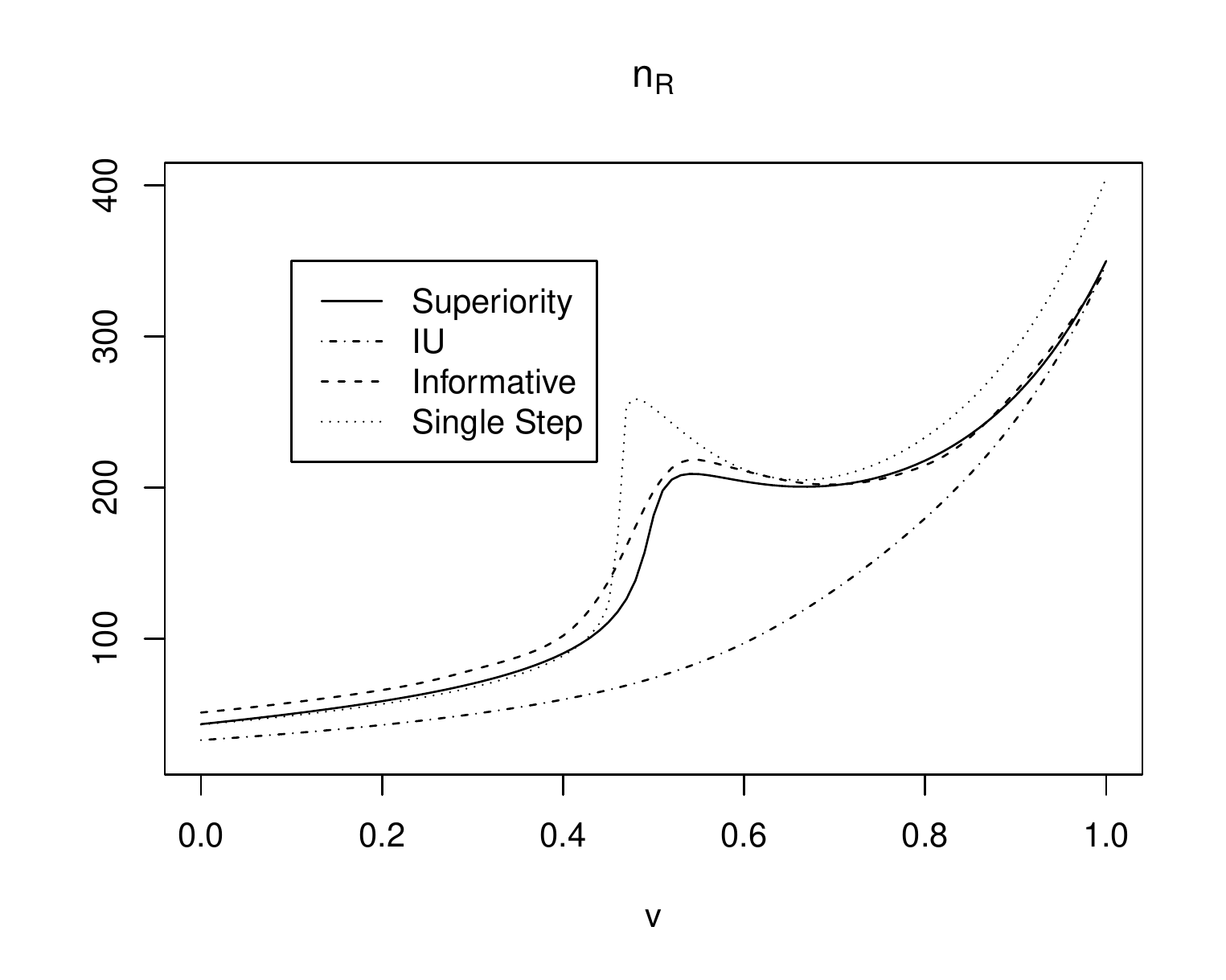}~\includegraphics[width= .5\textwidth]{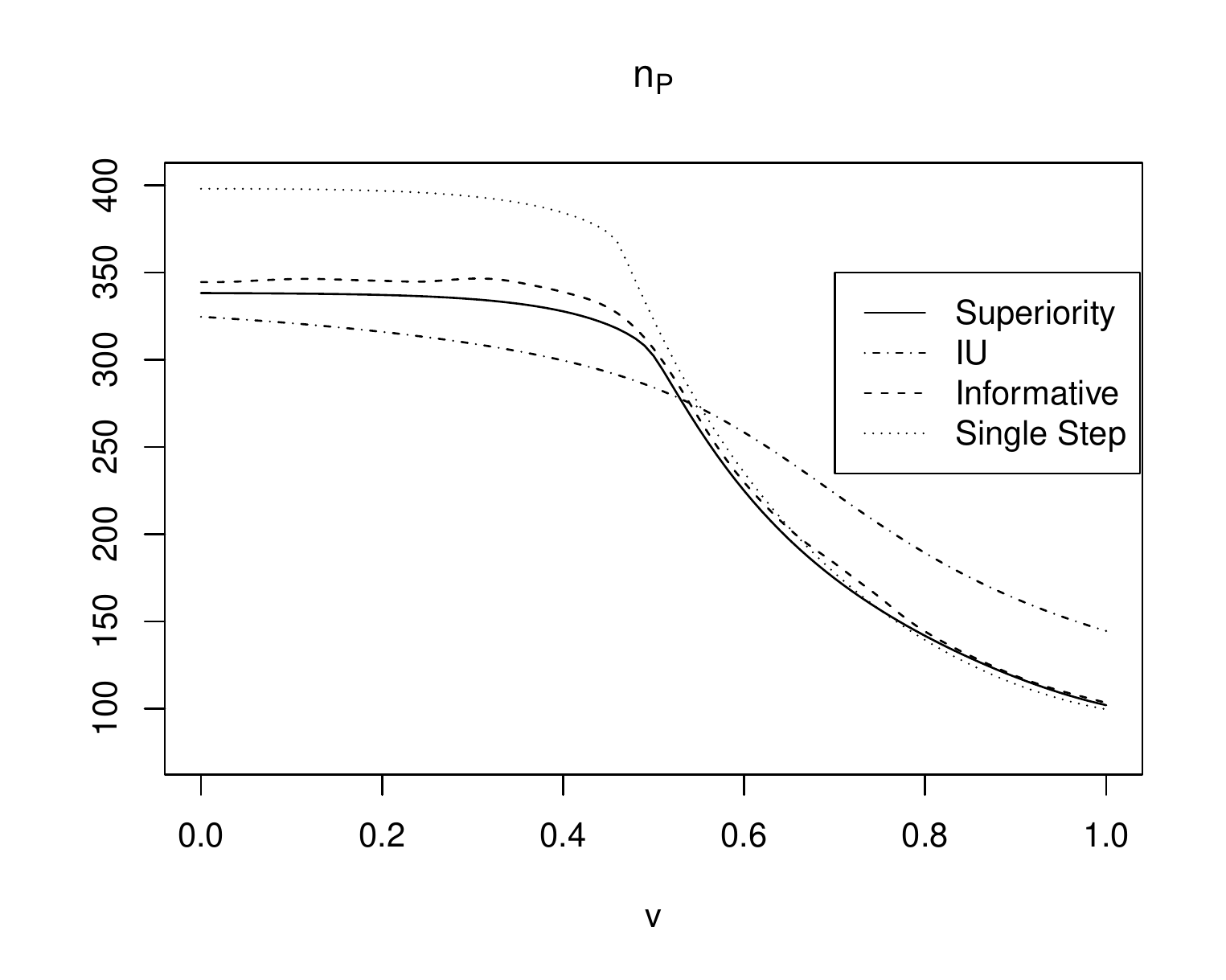}
\end{center}
\vspace*{-5mm}
\caption{\label{fig_samplesize_overv} Optimal sample sizes for the flexible non-inferiority design comparing the SCIs introduced in Section \ref{sec_SCIS} to the procedure of \citealp{BS15} without SCIs (see text for details). The $x$-axis is the ratio $v=$ reference effect / historical reference effect. Because of numerical instabilities, the values for the informative SCIs were smoothed.}
\end{figure}

Table~\ref{tab_filter} and Figure~\ref{fig_samplesize_overv} show the optimal sample sizes for the approach derived in \citealp{BS15} (``superiority filter'') and for the procedures involving SCIs as introduced in section~\ref{sec_SCIS}. As we can see, the informative SCIs require only slightly larger sample sizes than the procedure of \citealp{BS15} throughout all scenarios. The single step SCIs lead to the largest sample sizes for most values of $v$. For the scenario in which the reference presents with the full historical effect, the sample size needed for  the IU filter is higher than that needed for the informative SCIs. For the other scenarios, in which the reference effect is assumed smaller than the historical reference effect, the required sample size for the IU filter is smaller than that of the informative SCIs.

We can see in the lower left panel of Figure~\ref{fig_samplesize_overv}, that there is a ``bump'' in the sample size of the reference group for $v$ around 0.5 for the method without SCIs as well as the IU and informative SCIs. This is due to the fact, that in this region, the superiority filter, which is used for the interpretation of the study results with theses procedures, begins to declare ``$R>P$'', which directs the focus from testing $H_{EP}$ towards $H_{ER}$. Hence, a higher sample size in the reference group is beneficial in these situations.

Usually we are interested to keep the placebo group as small as possible. We can see from the lower right panel of Figure~\ref{fig_samplesize_overv} that the single step intervals require substantially larger placebo group sample sizes than the other procedures for small values $v$. For values of $v$ larger than 0.6 we observe, that the UI intervals require larger sample sizes in the placebo group than the other methods. The informative intervals seem to offer a good compromise with sample sizes near those of the procedure of \citealp{BS15}, while offering the benefit of SCIs.

\begin{figure}
\begin{center}
\includegraphics[width= .5\textwidth]{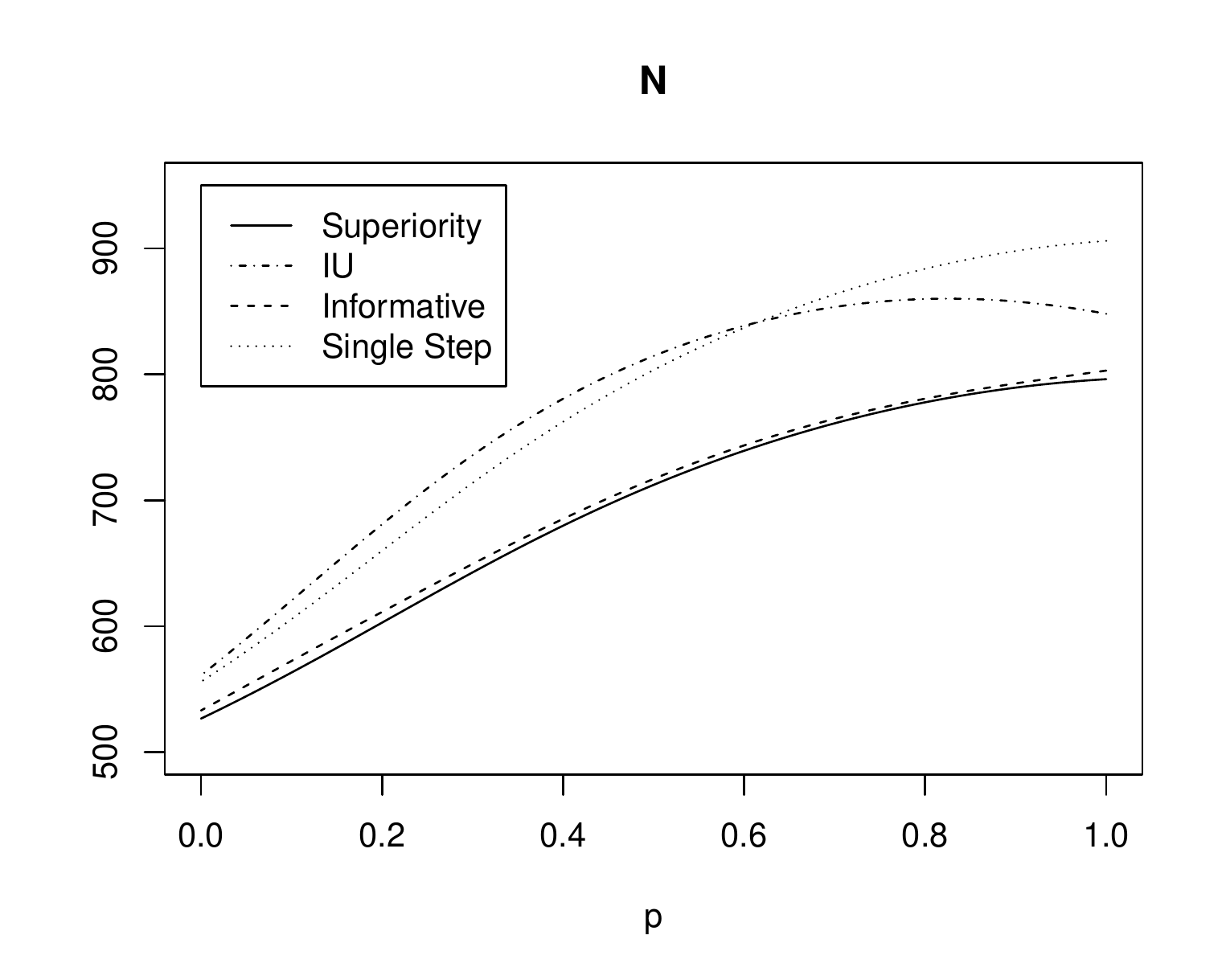}~\includegraphics[width= .5\textwidth]{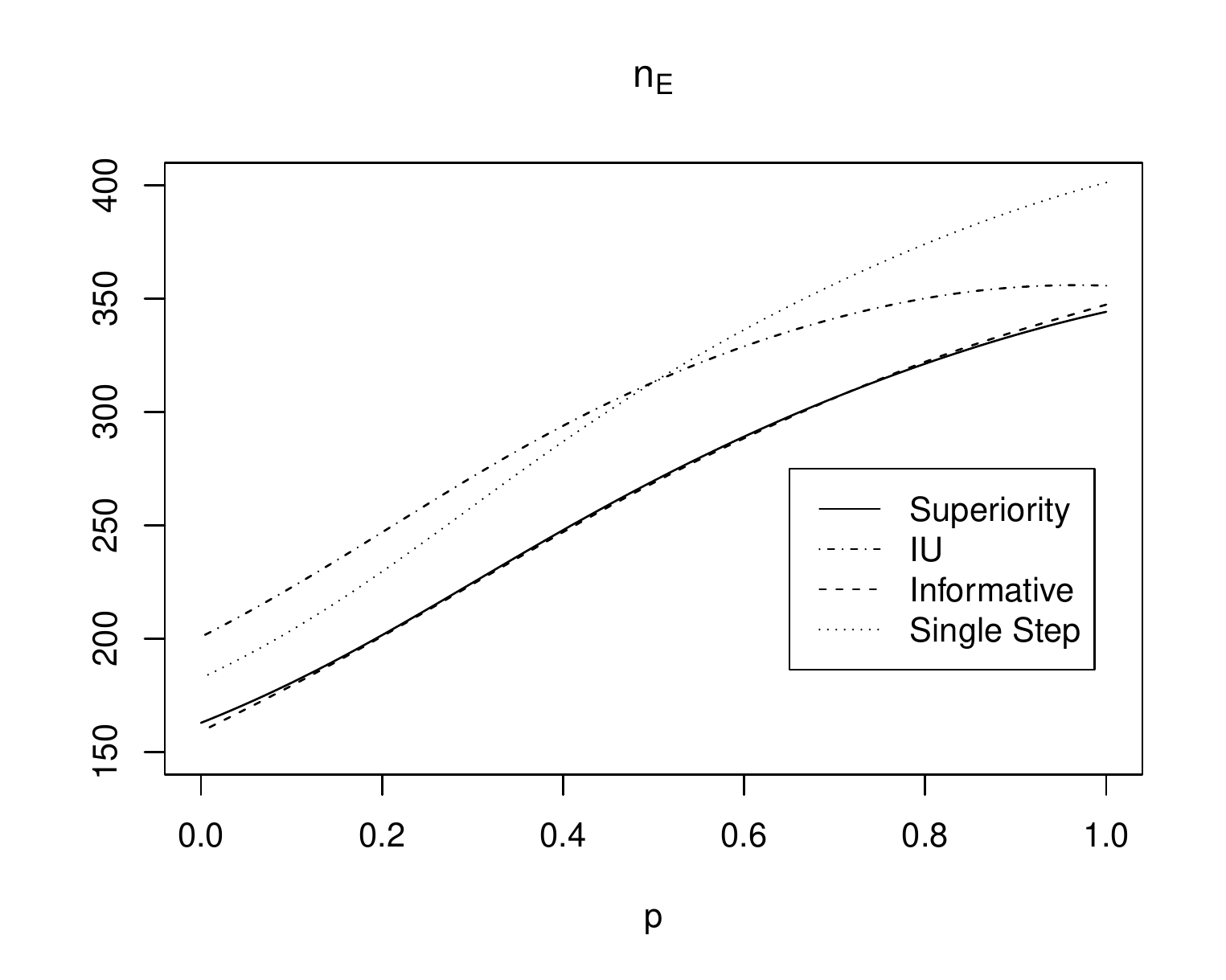}\\
\includegraphics[width= .5\textwidth]{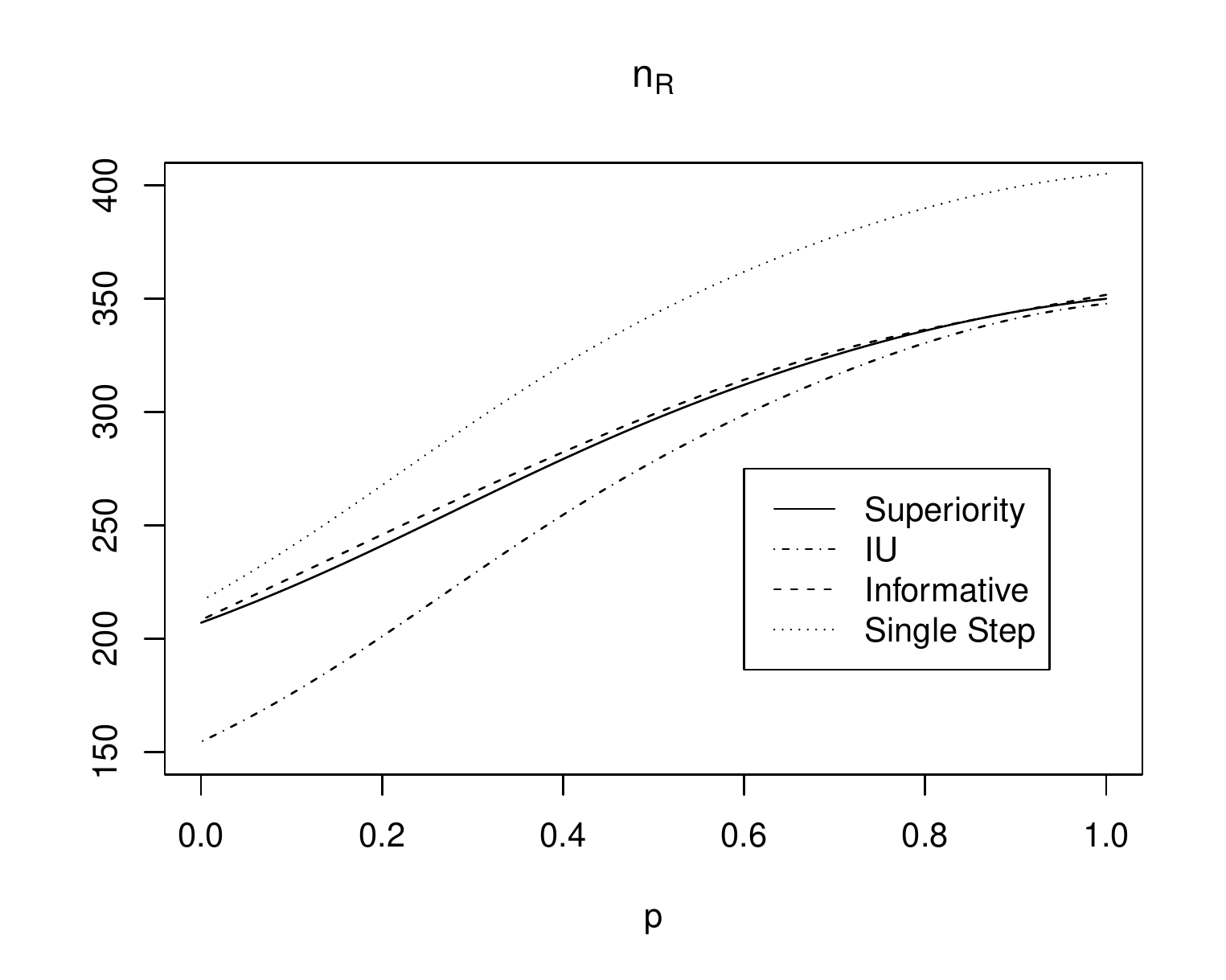}~\includegraphics[width= .5\textwidth]{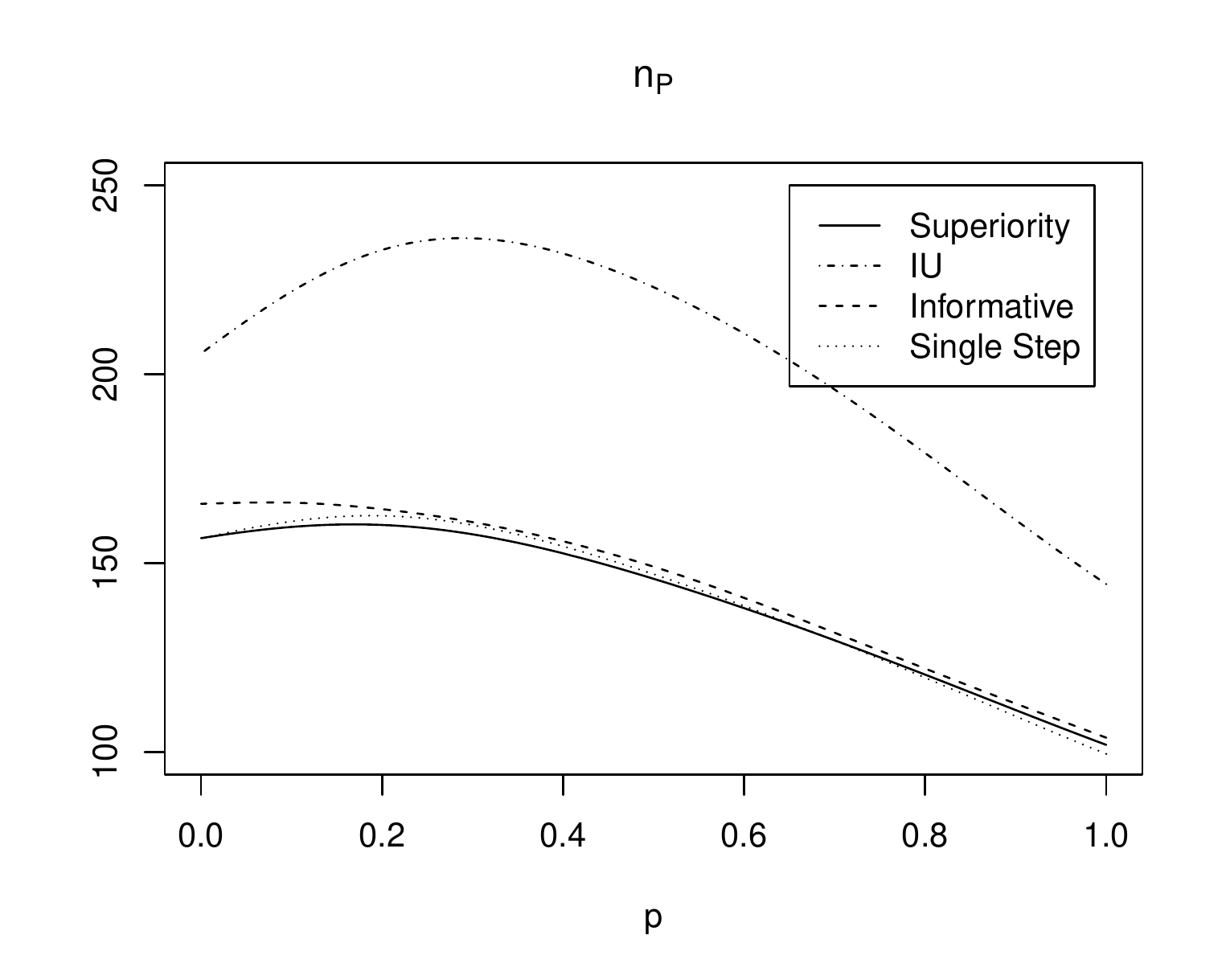}
\end{center}
\vspace*{-5mm}
\caption{\label{fig_samplesize_overp} Optimal sample sizes for the flexible non-inferiority design comparing the SCIs introduced in Section \ref{sec_SCIS} to the procedure of \citealp{BS15} without SCIs (see text for details). The $x$-axis is the probability $p$ that $v=1$. $v=2$ is assumed with probability $1-p$. Because of numerical instabilities, the values for the informative SCIs were smoothed.}
\end{figure}

As seen in Table~\ref{tab_filter} and Figure~\ref{fig_samplesize_overv}, the optimal sample size is very sensitive to the assumed reference effect in the current trial. Since there might be some uncertainty about the ratio $v$ of the reference effect in the study divided by the historical reference effect, \citealp{BS15} proposed an ``hybrid'' Bayesian-frequentist approach to the sample size calculation. In this approach, several ratios for $v$ are considered and weighted with an assumed probability of their occurrence. More general, denote the density of the probability distribution of $v$ on $[0,1]$ with $f(v)$. Then the weighted success probability is given by 
\begin{equation} \label{success_general}
S = \int_0^1 S_{v}f(v)dv,
\end{equation}
where $S$ is the targeted success probability and $S_v$~is the success probability if $v$ is the true ratio. \citealp{BS15} state that $S$ is often referred to as ``probability of success'' (cf.~\citealp{Kunzmann21}) or ``assurance'' (cf.~\citealp{Stallard09}). A very simple application of this approach is to define a grid of values for $v$ and assign probabilities for each of the values. To illustrate this approach we use a simple example: Since we do not expect the reference to fail completely, we expect $v=1$ with some probability~$p$ and we assume with probability $1-p$ that $v=3/4$. The probability of success in \eqref{success_general}, which is targeted at~$90\%$, as before, is then given by
\begin{equation} \label{success}
S = p S_{1} + (1-p) S_{3/4}
\end{equation}
Figure~\ref{fig_samplesize_overp} shows the optimal sample sizes obtained with this approach for $p\in [0,1]$. We see that for all values of $p$ the informative intervals yield sample sizes close to those of the procedure of \citealp{BS15}. We can also see that the IU intervals yield to a substantially larger placebo group, compared to the other methods.

\citealp{BS15} indicate, that ``in practical applications, the choice of the values for $v$ and $p$ should be driven by knowledge from prior studies. I.e. the values chosen for $v$ should reflect true ratios observed in prior studies and $p$ should be driven by the believes as to how likely the respective values for $v$ will be observed in the planned trial.''

\subsection{Example}\label{ex_interpretation}
To illustrate the simultaneous confidence intervals proposed in sections~\ref{sec_design} to \ref{sec_singlestep}, as well as their interpretation on our notion of ``success'', we give an example of different outcomes. We assume the setting of scenario~1 of the previous section, resulting in the following optimal sample sizes for the IU filter: $n_E=356$, $n_R=348$, $n_P=145$.

Without loss of generality, the observed placebo mean is $X_P=0$. Consider Table~\ref{tab_example}, where the lower confidence bounds $L_{EP}$ and $L_{ER}$ are calculated via~\eqref{lep} and~\eqref{ler} for different observed means $X_E$ and $X_R$. Furthermore, it is indicated if the filter~\eqref{filterE} is satisfied. Note, that with the given assumptions, ``$R>P$" can be stated if and only if $X_R>0.591$. If this filter is satisfied, $L_{ER}$ is of major interest and $L_{EP}$ is reported only additionally (in the table in parentheses). If the filter is not satisfied, then $L_{EP}$ is of major interest. In Tables~\ref{tab_example2} and \ref{tab_example3} the informative and single-step confidence bounds are calculated. Furthermore, it is indicated if the filter~\eqref{supfilter} is satisfied. With the given assumptions, ``$R>P$" can be stated if and only if $X_R>0.387$. For the informative SCIs $q=0.01$ was chosen. For more information on the choice of $q$ see section~\ref{sec_pos}.

In the first rows of Table~\ref{tab_example}, the observed mean in the experimental group equals the historical reference effect. In the first row, $X_R>0.591$, hence the IU filter is satisfied and the focus is on the bond $L_{ER}=\ell_{ER}$. Success is given via the column named Success~ER in the table, namely if $L_{ER}\ge-\delta_0$, which is the case here. In the second row, the IU filter is not satisfied, therefore, success is given by $L_{EP}=\ell_{EP}$ and is defined via Success~EP, namely that $L_{EP}\ge \delta_1$. One has proven that the new treatment has a~relevant effect, although the reference is only half as strong as historically. In the fourth example, the filter is also not satisfied. Now the effect of the experimental treatment is lower, therefore no success can be attested. This rule is stricter than the simple Koch-R\"ohmel design, where non-inferiority of $E$ versus~$R$ would be sufficient. Here one can even show superiority of $E$ versus~$R$ with $\ell_{ER}$, however, this has no merit because the reference is weak and $L_{ER}$ is only a shifted version of $L_{EP}$ and thus does not give additional information.

In the first two rows of Tables~\ref{tab_example2} and \ref{tab_example3} the superiority filter assesses ``$R>P$" and the respective confidence bounds $L_{ER}^{inf}$ and $L_{ER}^{S}$ are both larger than $-\delta_0$ leading to a successful outcome (Success~ER). In the last two rows, the observed mean $X_R$ is fairly small, leading the superiority filter to declare ``$R\leq P$". For the third row, $L_{EP}^{inf}$ and $L_{EP}^{S}$ are both larger than $\delta_1=0.5$ leading to a Success~EP. In the last row, $L_{EP}^{inf}$ and $L_{EP}^{S}$ both fall below that threshold, resulting in no claim of success. While both, the informative and the single-step confidence intervals come to the same conclusions, we can see that they differ in their size. In this example we observe that we always have $L_{EP}^{inf}>L_{EP}^{S}$. This is particularly useful in the third row, because superiority over placebo is the main conclusion here. Note that there is no such trend for the bounds for the comparison of E and R.

\begin{table}
\caption{Example for the determination of the SCI with different filters and the resulting success}
\begin{subtable}{\textwidth}
\caption{\label{tab_example} IU filter with corresponding SCIs. Success~ER is satisfied if the IU filter holds and $L_{ER}\ge -0.5$. Success~EP is satisfied if the IU filter does not hold and $L_{EP} > 0.5$. See text for further explanations.}\vspace{2mm}
\centering
\begin{tabular}{rrcrrrrcc}
  \hline
$X_E$ & $X_R$ & IU filter & $\ell_{EP}$ & $\ell_{ER}$ & $L_{EP}$ & $L_{ER}$ & Success EP & Success ER \\ 
  \hline
 1.00 & 1.00 & yes & 0.614 & -0.295 & (0.205) & -0.295~ & no  &yes\\ 
 1.00 & 0.50 & no & 0.614 & 0.205 & 0.614~ & (0.114) 		& yes	&no  \\ 
 1.00 & 0.30 & no & 0.614 & 0.404 & 0.614~ & (0.114) 		& yes	&no  \\
 0.80 & 0.30 & no & 0.414 & 0.205 & 0.414~ & (-0.086) 	& no 	&no \\ 
   \hline
\end{tabular}
\end{subtable}\\[4mm]

\begin{subtable}{\textwidth}
\caption{\label{tab_example2} Superiority filter with informative SCIs ($q=0.01$). Success~ER is satisfied if the superiority filter holds and $L^{inf}_{ER}\ge -0.5$. Success~EP is satisfied if the superiority filter does not hold and $L^{inf}_{EP} > 0.5$. See text for further explanations.}\vspace{2mm}
\centering
\begin{tabular}{rrcrrcc}
  \hline
$X_E$ & $X_R$ & superiority filter & $L^{inf}_{EP}$ & $L^{inf}_{ER}$ & Success EP & Success ER \\ 
  \hline
 1.00 & 1.00 & yes & 0.561 & -0.340 & no  & yes\\ 
 1.00 & 0.50 & yes & 0.607 & 0.063 	& no  & yes\\ 
 1.00 & 0.30 & no & 0.611 & 0.228 		& yes	& no  \\ 
 0.80 & 0.30 & no & 0.407 & 0.063 		& no 	& no \\  
   \hline
\end{tabular}
\end{subtable}\\[4mm]

\begin{subtable}{\textwidth}
\caption{\label{tab_example3} Superiority filter and single-step SCIs. Success~ER is satisfied if the superiority filter holds and $L^S_{ER}\ge -0.5$. Success~EP is satisfied if the superiority filter does not hold and $L^S_{EP} > 0.5$. See text for further explanations.}\vspace{2mm}
\centering
\begin{tabular}{rrcrrcc}
  \hline
$X_E$ & $X_R$ & superiority filter & $L^{S}_{EP}$ & $L^{S}_{ER}$ & Success EP & Success ER \\ 
  \hline
  1.00 & 1.00 & yes & 0.560 & -0.337 & no  & yes\\ 
 1.00 & 0.50 & yes & 0.560 & 0.163 	& no  & yes\\ 
 1.00 & 0.30 & no & 0.560 & 0.363 		& yes	& no  \\ 
 0.80 & 0.30 & no & 0.360 & 0.163 		& no 	& no \\  
   \hline
\end{tabular}
\end{subtable}
\end{table}

\subsection{Probability of Success}\label{sec_pos}
We now compare the probability of success between the SCIs presented in this paper and the method of \citealp{BS15} using the superiority filter \eqref{supfilter} but without calculating SCIs. With the comparison to the latter procedure we can assess the ``cost" (in terms of probability of success) for the additional information the SCIs provide. We ran a simulation with 100,000 repetitions of observations in the settings of scenarios 1 and 2 of section~\ref{sec_optimalN}. We counted how often the filters were satisfied and, in each case, how often the relevant success occurred (i.e., proof of non-inferiority if the filter is valid, proof of $\delta_1$-superiority if the filter is not valid) and what the median of the confidence bounds is. Note, that the mean of $L_{ER}$ cannot be reported because this bound is minus infinity in case the gatekeeper (superiority of $E$ versus~$P$) fails to be shown. The mean of $L_{EP}$, however, did not differ strongly from its median, so we show here only the median as a~measure of central tendency of the confidence bounds. However, failure of the gatekeeper was almost never a~problem. For the informative SCIs we investigated different choices for $q$. Here, we only present the results for $q=0.01$, since they yielded the best results. 

\begin{figure}
\begin{center}
\includegraphics[width= .5\textwidth]{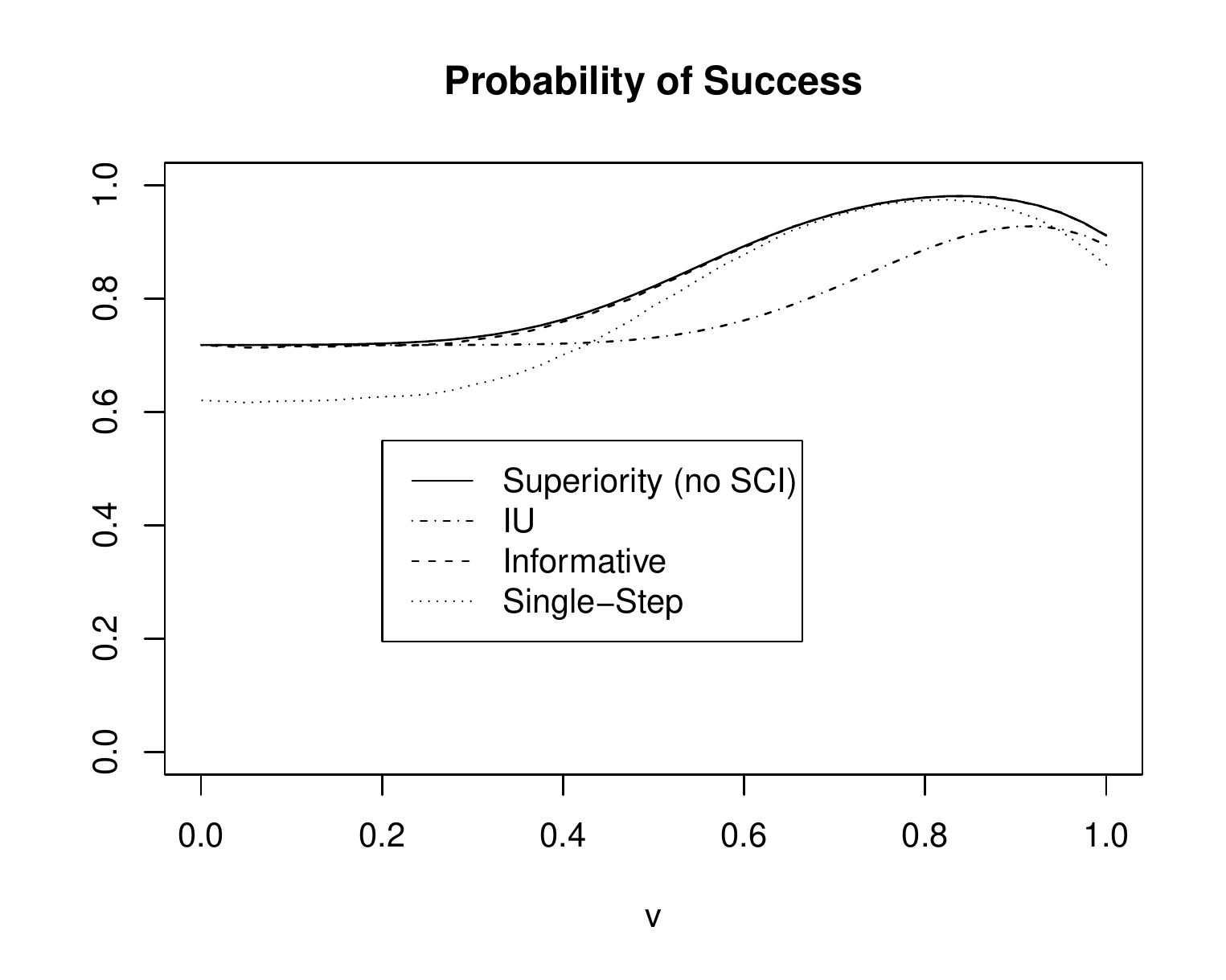}~\includegraphics[width= .5\textwidth]{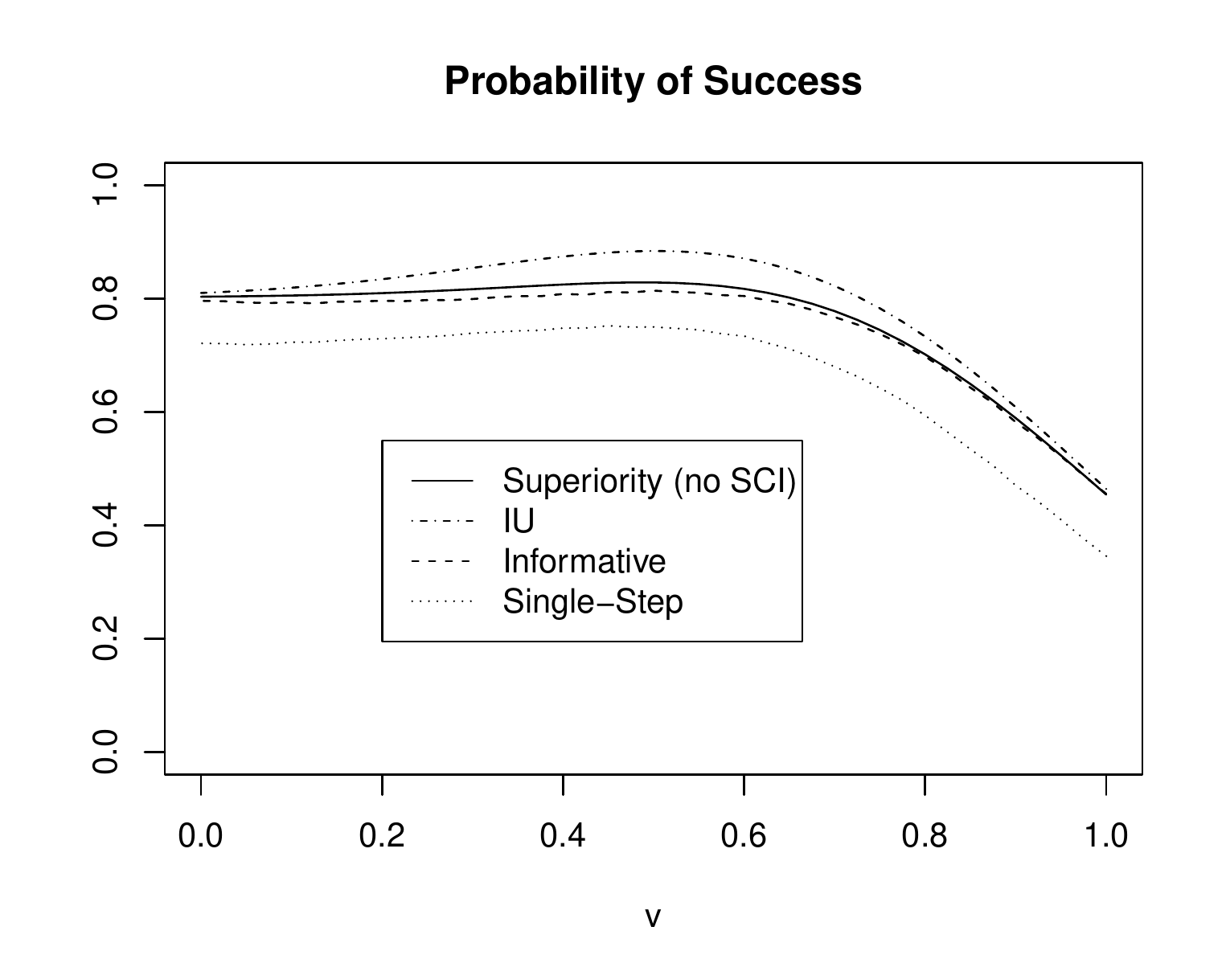}\\
\includegraphics[width= .5\textwidth]{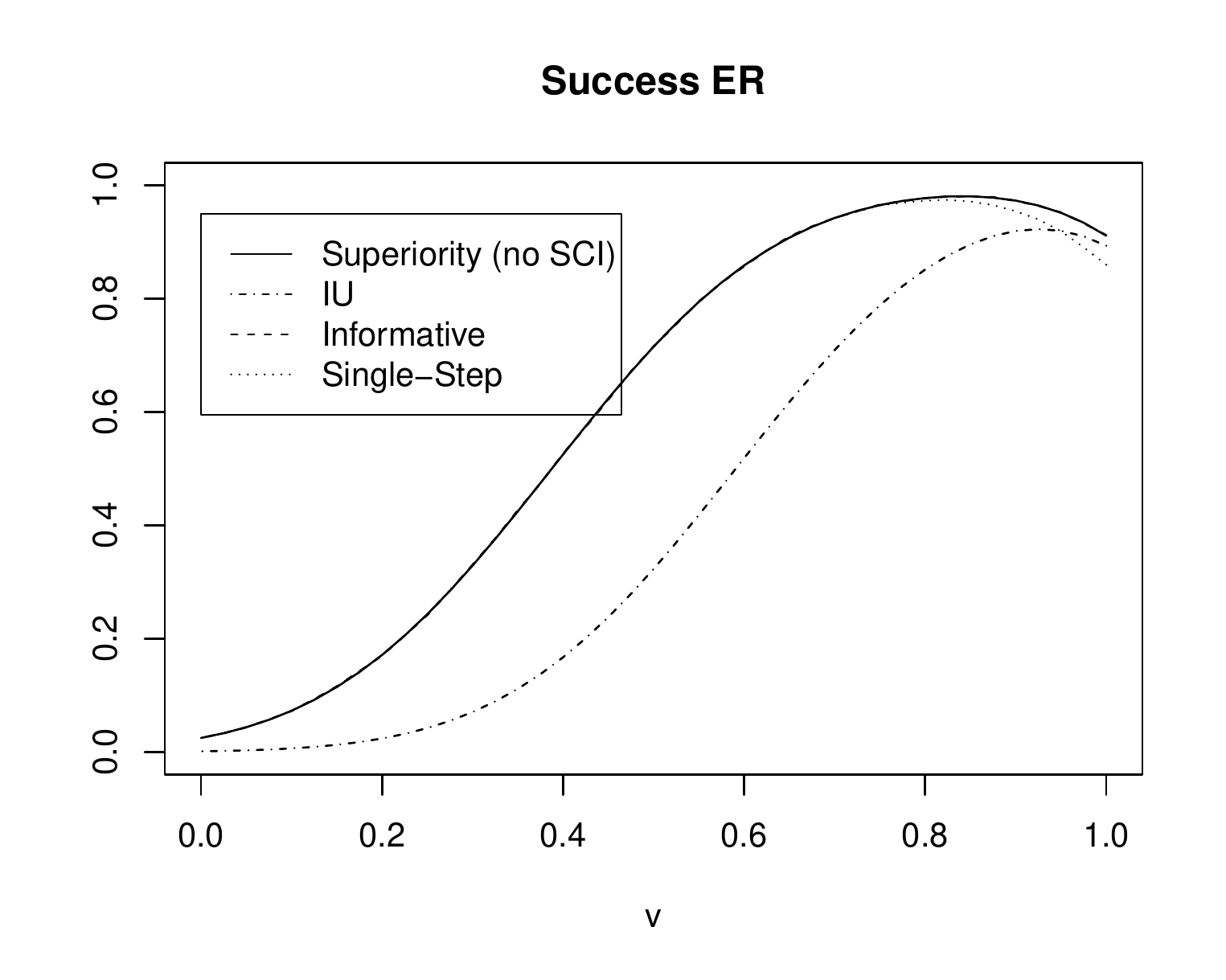}~\includegraphics[width= .5\textwidth]{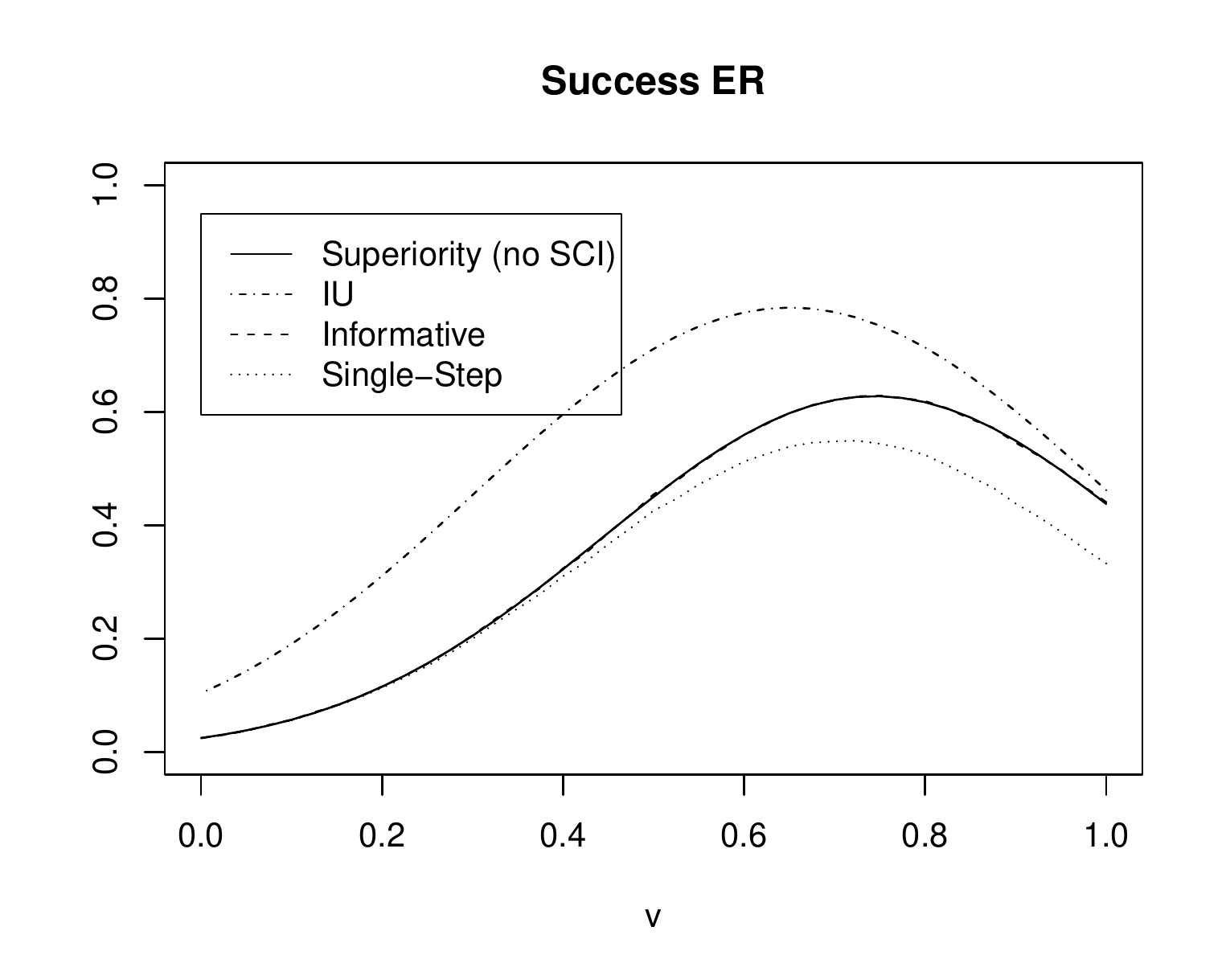}\\
\includegraphics[width= .5\textwidth]{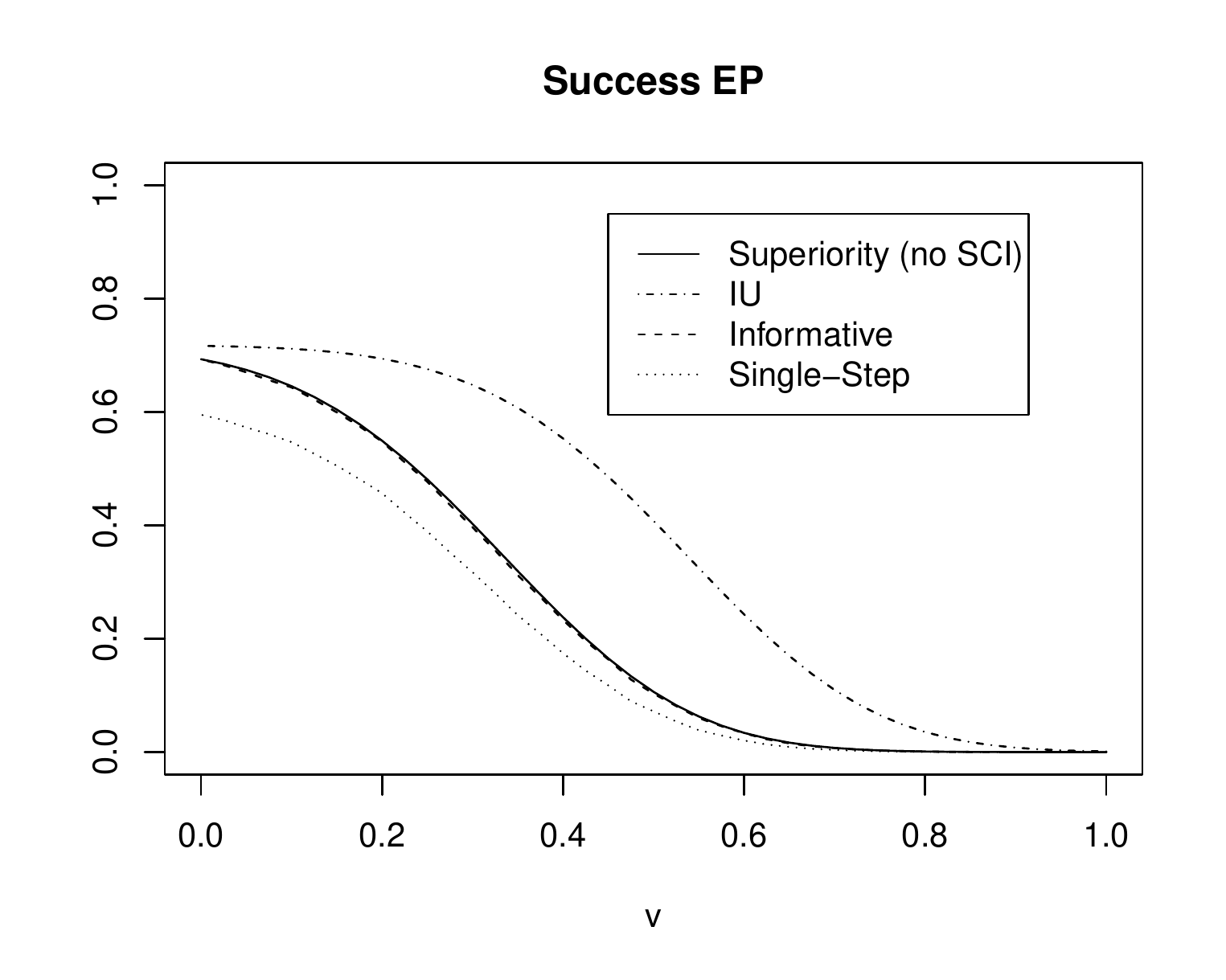}~\includegraphics[width= .5\textwidth]{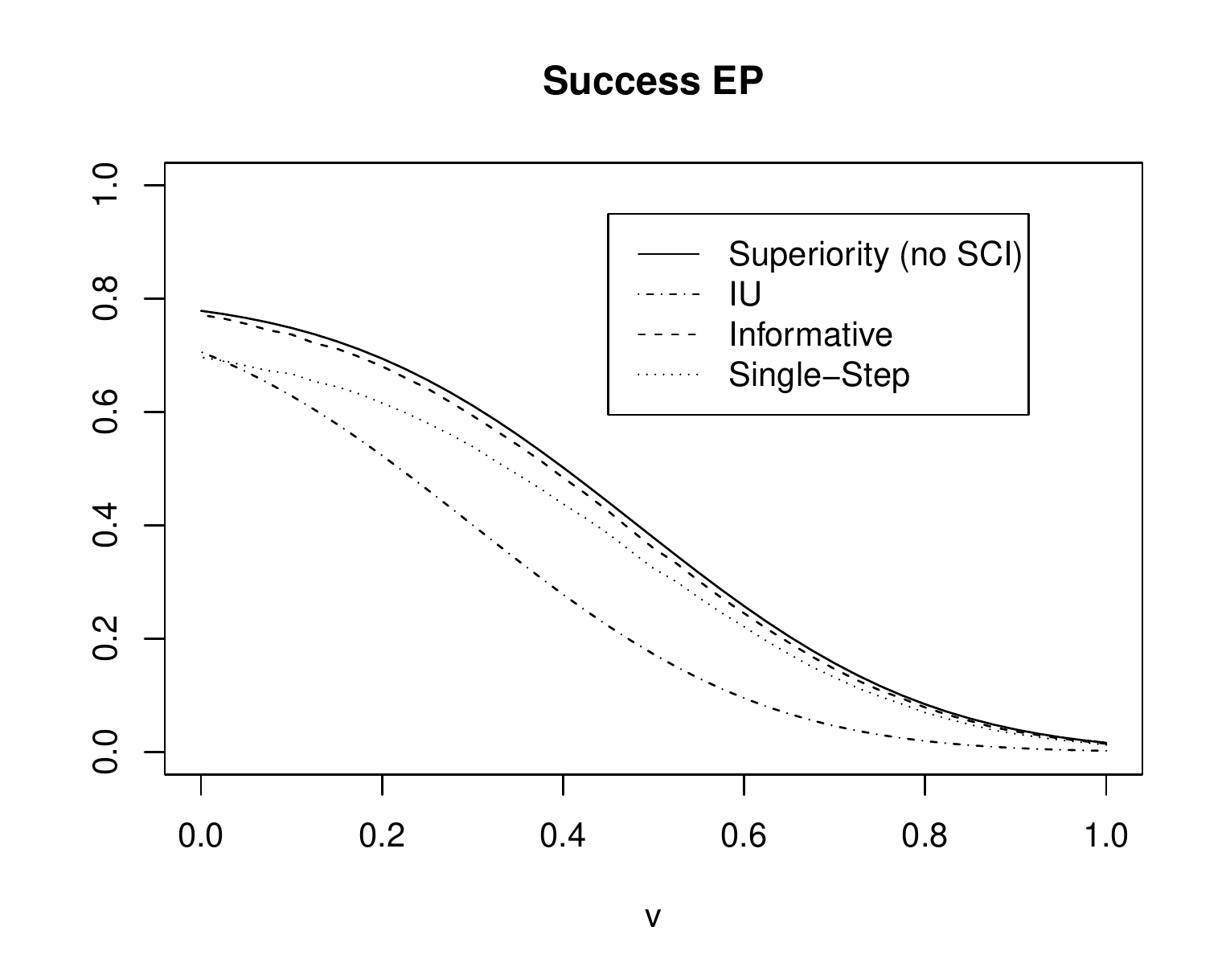}\\
\end{center}
\vspace*{-5mm}
\caption{\label{fig_power} Probability of Success, success with valid filter (Power ER) and success without valid filter (Power EP) from simulations in scenario 1 (left column) and scenario 2 (right column). The $x$-axis is the ratio $v=$ reference effect / historical reference effect.}
\end{figure}

\begin{figure}
\begin{center}
\includegraphics[width= .5\textwidth]{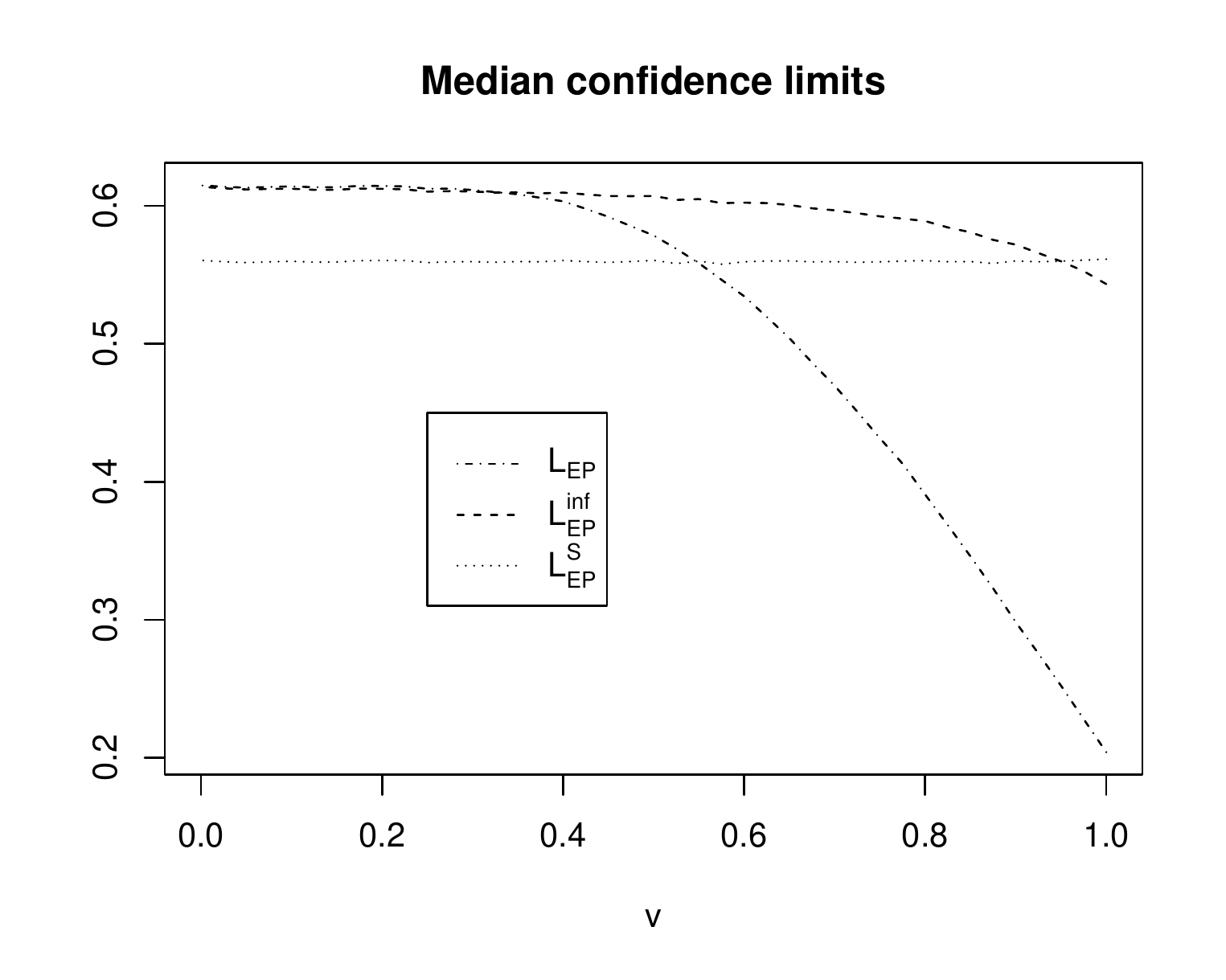}~\includegraphics[width= .5\textwidth]{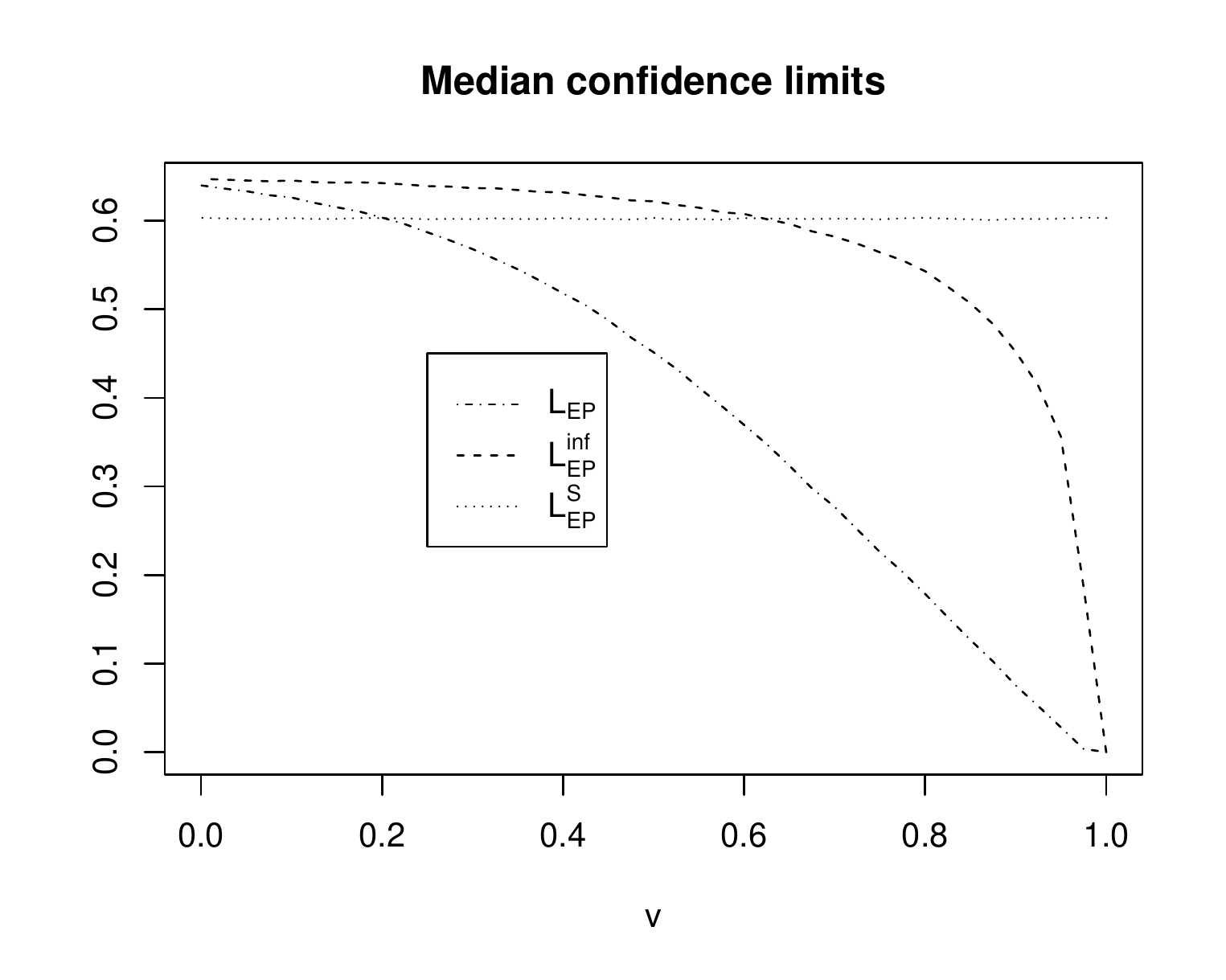}\\
\includegraphics[width= .5\textwidth]{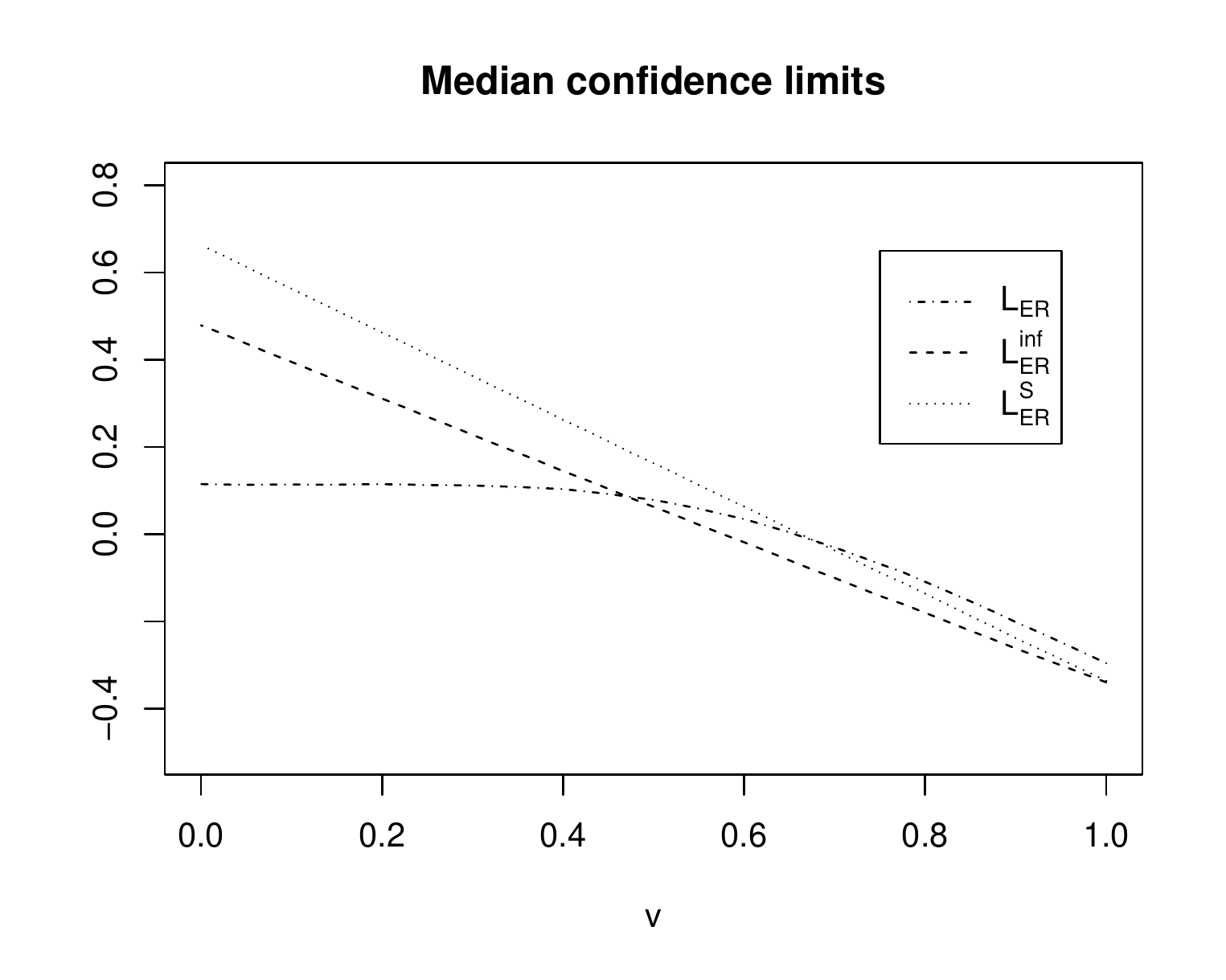}~\includegraphics[width= .5\textwidth]{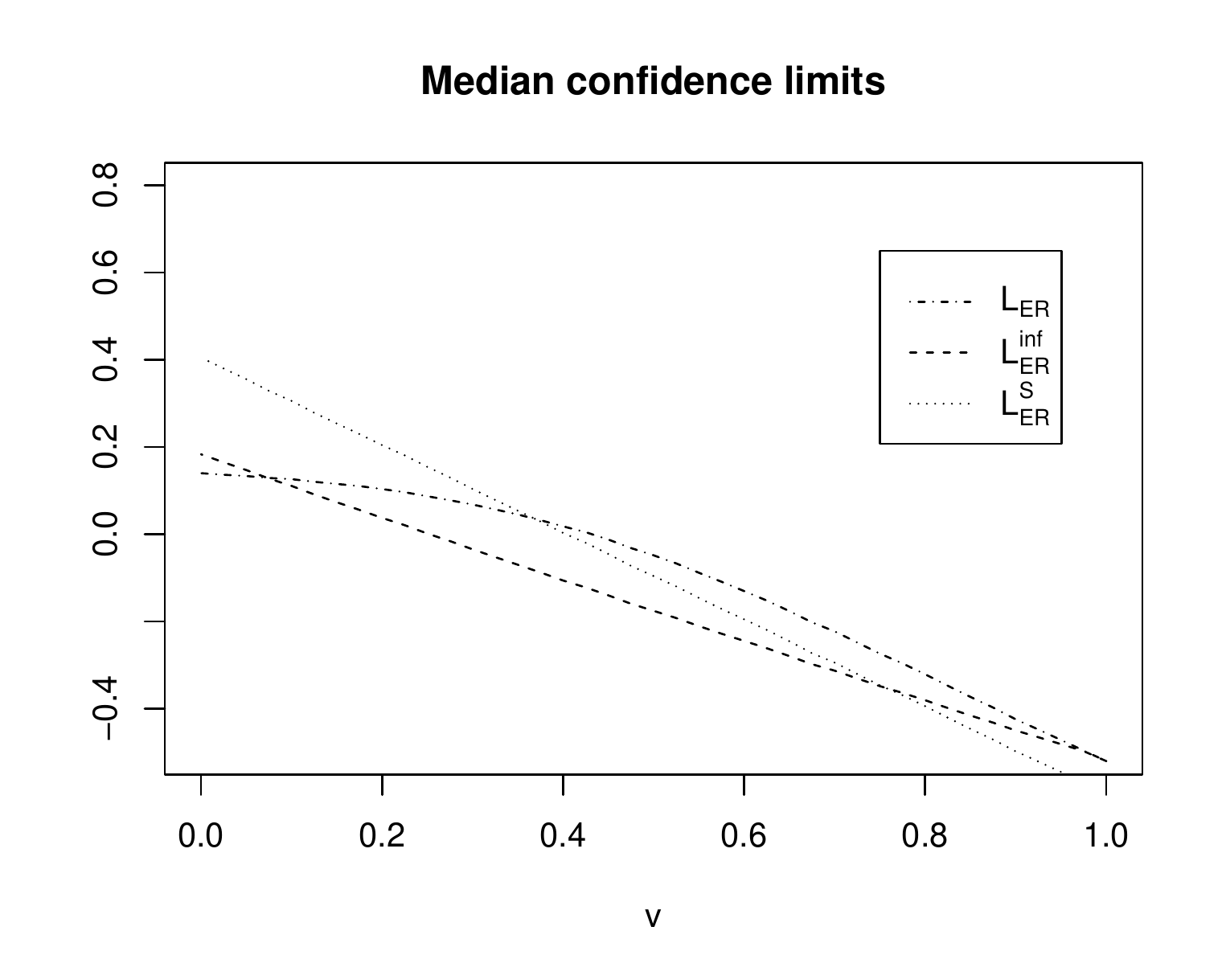}\\
\end{center}
\vspace*{-5mm}
\caption{\label{fig_limits} Median lower confidence limits from simulations in scenario 1 (left column) and scenario 2 (right column). The $x$-axis is the ratio $v=$ reference effect / historical reference effect.}
\end{figure}

The results of the simulation are presented in table~\ref{tab_characteristics} and figures~\ref{fig_power} and \ref{fig_limits}. We see that there is only very little loss in probability of success when comparing the informative SCIs interpreted with the help of the superiority filter, to the procedure of \citealp{BS15} using the superiority filter without SCIs. The maximum observed loss in probability of success observed in our simulations was about $1\%$, which is quite low considering the benefit in interpretation the SCIs offer. Throughout all simulations we observed that the single-step confidence intervals interpreted with the help of the superiority filter yielded a lower probability of success than the procedure of \citealp{BS15} or the informative SCIs. The loss in probability of success compared to using the informative SCIs amounted up to approx. $10\%$ in some scenarios. Comparing the SCIs with IU filter to the procedures using the superiority filter, we see that in scenario 1, where we the sample size is optimized under the assumption that the reference is as strong as historically observed, the IU filter offers substantial less probability of success for most reference effects (ratio of $v$=reference effect/historical reference effect $>0.2$). The loss in probability of success was as high as $13\%$ for $v=0.65$. In scenario 2 however, where the sample size is optimized under the assumption, that the reference will only show half it's historical effect in the present trial, the SCIs with IU filter show best results in terms of probability of success. However, the observed loss for the superiority filter is at most around $6\%$, which is smaller than the loss the IU filter procedure presented with in scenario 1. 

\begin{table}
\caption{\label{tab_characteristics} Characteristics of simultaneous confidence intervals for the settings of scenarios 1 and 2 of section~\ref{sec_optimalN} for different ratios of $v$=reference effect/historical reference effect. Without SCI is the probability of success for the procedure introduced in \citealp{BS15}, using the superiority filter but no SCIs. For the informative and single-step intervals, the superiority filter was used for interpretation of study results.}
\begin{subtable}{\textwidth}
\caption{\label{subtab_char_filter5} Scenario 1 (Sample size planning with $v=1$). Sample size: $N=n_E+n_R+n_P=356+348+145=849$.}\vspace{2mm}
\centering
\begin{tabular}{r|rr|rrrr}
  \hline
	& \multicolumn{2}{|c|}{Filter satisfied} & \multicolumn{4}{c}{Probability of Success}
	\\
	\hline
 v & IU & superiority & Without SCI & IU & Informative & Single-Step \\
\hline
 1.00 & 98.1 & 99.9 & 91.2 & 89.5 & 91.2 & 86.0 \\
 0.75 & 79.2 & 96.7 & 96.9 & 85.5 & 96.9 & 96.6 \\
 0.50 & 32.7 & 71.6 & 82.2 & 73.2 & 81.9 & 78.8 \\
 0.25 & 4.3  & 24.2 & 72.4 & 71.7 & 71.9 & 63.1 \\
 0.00 & 0.1  & 2.5  & 72.0 & 72.0 & 71.8 & 62.1 \\
   \hline
\end{tabular}
\end{subtable}\\[4mm]

\begin{subtable}{\textwidth}
\caption{\label{subtab_char_filter1} Scenario 2 (Sample size planning with $v=0.5$). Sample size: $N=n_E+n_R+n_P=227+75+285=587$.}\vspace{2mm}
\centering
\begin{tabular}{r|rr|rrrr}
  \hline
	& \multicolumn{2}{|c|}{Filter satisfied} & \multicolumn{4}{c}{Probability of Success}
	\\
	\hline
 v & IU & superiority & Without SCI & IU & Informative & Single-Step \\
\hline
 1.00 & 99.5 & 97.1 & 45.7 & 46.8 & 45.6 & 34.6 \\
 0.75 & 95.0 & 82.4 & 74.6 & 78.4 & 73.8 & 64.2 \\
 0.50 & 75.0 & 49.0 & 83.0 & 88.4 & 81.4 & 75.0 \\
 0.25 & 38.4 & 15.9 & 81.3 & 84.5 & 79.8 & 73.3 \\
 0.00 & 10.5 & 2.5  & 80.6 & 81.2 & 79.6 & 72.1 \\
   \hline
\end{tabular}
\end{subtable}

\end{table}

\section{Example -- Application to trial data}\label{sec_example}
Next, we apply the introduced confidence intervals to data of a three-arm study on major depressive disorder by \citealp{Higu11}, which is also discussed in \citealp{HT11} and \citealp{BS15}. We will first show the example in it's original version, as shown in \citealp{Higu11} as well as the application of the adaptive test strategy of \citealp{BS15}. After that, we will apply the IU and the informative ($q=0.01$) confidence intervals to the same data.

\citealp{Higu11} investigated the efficacy and safety of 6-week treatment with duloxetine (E) to those of paroxetine (R) and placebo (P) in a double-blinded, randomized, active controlled, parallel-group study. It's primary endpoint was the HAM-D17 change from baseline at 6 weeks and the statistical analysis was planned to test the superiority of duloxetine over placebo and the non-inferiority of duloxetine compared to paroxetine in hierarchical order. The non-inferiority margin was set as $\delta_0=2.5$. The following mean decreases in the HAM-D17 were observed: $10.2\pm 6.1$ (mean$\pm$sd) in the duloxetine group, $9.4\pm 6.9$ in the paroxetine group and $8.3\pm 5.8$ in the placebo group with sample sizes $n_E=147,\ n_R=148,\ n_P=145$. 

The resulting (unadjusted) 97.5\% lower confidence interval for the difference in means between duloxetine and placebo is in this case given by (0.53,\ $\infty$) (i.e.\ $\ell_{EP}= 0.53$), indicating superiority of duloxetine over placebo. The (unadjusted) 97.5\% lower confidence interval for the difference in means between duloxetine and paroxetine can be calculated to be (-0.69,\ $\infty$) (i.e.\ $\ell_{ER}= -0.69$) and excludes $-\delta_0=-2.5$ indicating non-inferiority of duloxetine compared to paroxetine. However, the superiority of paroxetine over placebo could not be established, because the lower 97.5\% confidence interval for this comparison is given by (-0.37,\ $\infty$) and therefore includes 0. \citealp{Higu11} concluded that non-inferiority of duloxetine compared to paroxetine did not have assay sensitivity.

\citealp{BS15} applied their adaptive test strategy to the data, employing the superiority filter \eqref{supfilter} to determine, whether paroxetine was sufficiently efficacious. They used $\delta_1=2.5$ for the investigation of superiority of E vs. P. As seen above duloxetine can be shown to be superior to placebo (i.e.\ $H_{EP}^S$ is rejected). However, using the superiority filter \eqref{supfilter}, ``$R>P$'' can not be concluded, which is why as a next step they tested for $\delta_1$-superiority of duloxetine over placebo ($H_{EP}^{\delta_1}$). Since the lower 97.5\% confidence interval for this comparison includes $\delta_1=2.5$, $H_{EP}^\delta$ could not be rejected, and the study was not successful in the sense that there was not enough evidence to declare duloxetine sufficiently efficacious.

For illustrative purposes, \citealp{BS15} investigated how their procedure would perform, if the observed mean change in the duloxetine group would have been $12.2$ with all other parameters unchanged. Under this assumption, duloxetine can be shown to be superior to placebo (i.e.\ $H_{EP}^S$ is rejected) and the superiority filter still concludes ``$R\leq P$''. However, one now has $\ell_{EP}=2.53$, leading to a claim of $\delta_1$ superiority of duloxetine to placebo, which in case can be interpreted as a successful study. \citealp{BS15} however note that the confidence intervals reported ``while appropriate for deriving the test decision in the hierarchical test, are no simultaneous confidence intervals, and therefore do not have simultaneous coverage probability''.

We now want to apply the simultaneous confidence intervals introduced in this manuscript to the original data of this study. Applying the intuitive approach outlined in figure~\ref{fig_design_intuitive}, we note that $H_{EP}: \mu_E-\mu_P\leq 0$ can be rejected. The IU and the superiority filters both conclude ``$R\leq P$'' and the IU simultaneous confidence bounds are given as: $L_{EP}=\ell_{EP}=0.53$ and $L_{ER}=\ell_{EP}-\delta_0=-1.97$. Since $L_{EP}$ covers the superiority margin $\delta_1=2.5$ we cannot establish ``success'' of the study. The informative confidence bounds are gives as: $L^{inf}_{EP}=0.528$ and $L^{inf}_{ER}=-1.67$. Since $L^{inf}_{EP}$ also covers the superiority margin $\delta_1=2.5$ we cannot establish ``success'' of the study with the informative SCIs either. However, the simultaneous confidence bounds $L_{EP}$ and $L_{ER}$ resp. $L^{inf}_{EP}$ and $L^{inf}_{ER}$ allow for an interpretation on the magnitude of treatment effects, which was not possible before.

Applying the confidence intervals to the hypothetical case where we assume that the observed mean change in the duloxetine group is $12.2$ with all other parameters unchanged, we obtain $\ell_{EP}=2.53$ and $\ell_{ER}=-0.69$. Again, $H_{EP}: \mu_E-\mu_P\leq 0$ can be rejected and the IU and the superiority filter both conclude ``$R\leq P$''. The IU simultaneous confidence bounds are now given as: $L_{EP}=\ell_{EP}=2.53$ and $L_{ER}=\ell_{EP}-\delta_0=0.03$. Since $L_{EP}>\delta_1$ we now conclude $\delta_1$ superiority of duloxetine over placebo, and therefore a ``success'' of the study. For the informative SCIs we obtain $L^{inf}_{EP}=2.53$ and $L^{inf}_{ER}=-0.59$. Since $L^{inf}_{EP}>\delta_1$ we also conclude $\delta_1$ superiority of duloxetine over placebo, and therefore a ``success'' of the study with the informative SCIs. While \citealp{BS15} came to the same conclusions on the significance tests, the proposed confidence bounds offer the advantage, that we can derive conclusions on the size of the effects of duloxetine compared to placebo and paroxetine. 

\section{Summary and Discussion} \label{sec_discussion}
\citealp{BS15} introduced a flexible extension to the three-arm `gold-standard' non-inferiority design, in which it is possible to declare success of the study, even if the reference fails. We complemented the previous work with compatible simultaneous confidence intervals, thus providing more information than hypothesis testing alone. We introduced three methods to calculate confidence intervals in the given design: stepwise confidence intervals based on those introduced in \citealp{HB99}, ``informative'' SCIs based on the work of \citealp{SB14} and single-step confidence intervals. While the informative and the single-step intervals rely on a ``filter'' for the interpretation of study results, the intervals based on \citealp{HB99} offer an ``intrinsic'' filter (IU filter) guaranteeing FWER control for all sample sizes.

We did multiple simulations in which we compared the different procedures for calculating SCIs and also the procedure in \citealp{BS15} in different scenarios. We saw that substantial power losses may occur with the IU filter compared to the superiority filter. However, when the reference effect is anticipated to be smaller than historically observed (and the study is planned accordingly) the SCIs with IU filter might perform better than the superiority filter. We observed only very little loss in the probability of success for the informative SCIs compared to the procedure in \citealp{BS15} which does not yield SCIs at all. 

Finally, we applied the different SCIs to data from a real clinical trial, exemplifying the gain of calculating SCIs, which lies is the possibility to make statements of the size of the effects.

\bibliographystyle{apalike}
\bibliography{gold_sci} 

\begin{thebibliography}{}

\bibitem[Brannath et~al., 2022]{BS15}
Brannath, W., Scharpenberg, M., and Schmidt, S. (2022).
\newblock A flexible design for non-inferiority studies.
\newblock {\em Stat. Med.}, (10.1002/bimj.200510169).

\bibitem[EMA, 2010]{EMA2010}
EMA (2010).
\newblock Reflection paper on the need for active control in therapeutic areas
  where use of placebo is deemed ethical and one or more established medicines
  are available.
\newblock {\em European Medicines Agency, London}, (Draft, EMA/759784/2010).

\bibitem[Hauschke and Pigeot, 2005]{HP05}
Hauschke, D. and Pigeot, I. (2005).
\newblock Establishing efficacy of a new experimental treatment in the `gold
  standard' design.
\newblock {\em Biom. J.}, 47(6):782--786.

\bibitem[Hida and Tango, 2011]{HT11}
Hida, E. and Tango, T. (2011).
\newblock On the three-arm non-inferiority trial including a placebo with a
  prespecified margin.
\newblock {\em Stat Med}, 30:224--231.

\bibitem[Higuchi et~al., 2009]{Higu11}
Higuchi, T., Murasaki, M., and Kamijima, K. (2009).
\newblock Clinical evaluation of duloxetine in the treatment of major
  depressive disorder-placebo- and paroxetine-controlled double-blind
  comparative study.
\newblock {\em Jpn J Clin Psychopharmacol}, 12:1613--34.

\bibitem[Hsu, 1996]{H96}
Hsu, J. (1996).
\newblock {\em Multiple Compariosons -- Theory and methods}.
\newblock Chapman \& Hall/CRC.

\bibitem[Hsu and Berger, 1999]{HB99}
Hsu, J.~C. and Berger, R.~L. (1999).
\newblock Stepwise confidence intervals without multiplicity adjustment for
  dose-response and toxicity studies.
\newblock {\em J. Am. Stat. Ass.}, 94:468--482.

\bibitem[Kaiser et~al., 2022]{Kaiser22}
Kaiser, T., Volkmann, C., Karyotaki, E., Cuijpers, P., and Brakemeier, E.
  (2022).
\newblock Heterogeneity of treatment effects in trials on psychotherapy of
  depression.
\newblock {\em APA PsycArticles}, page Advance online publication.

\bibitem[Koch and R{\"o}hmel, 2004]{KR05}
Koch, A. and R{\"o}hmel, J. (2004).
\newblock Hypothesis testing in the ``gold standard'' design for proving the
  efficacy of an experimental treatment relative to placebo and a reference.
\newblock {\em J. Biopharm. Statist.}, 14(2):315--325.

\bibitem[Kunmann et~al., 2021]{Kunzmann21}
Kunmann, K., Grayling, M., Lee, K., Robertson, D., Rufibach, K., and Wason, J.
  (2021).
\newblock A review of bayesian perspectives on sample size derivation for
  confirmatory trials.
\newblock {\em The American Statistician}, 75(4):424--432.

\bibitem[Maurer et~al., 1995]{MHL95}
Maurer, W., Hothorn, L., and Lehmacher, W. (1995).
\newblock Multiple comparisons in drug clinical trials and preclinical assays:
  a-priori ordered hypotheses.
\newblock In {\em Biometrie in der chemisch-pharmazeutischen Industrie}, pages
  3--18. J.~Vollmar, Fischer Verlag, Stuttgart.

\bibitem[Schl{\"o}mer and Brannath, 2013]{SB13}
Schl{\"o}mer, P. and Brannath, W. (2013).
\newblock Group sequential designs for three-arm `gold standard'
  non-inferiority trials with fixed margin.
\newblock {\em Stat. Med.}, 32(28):4875--4889.

\bibitem[Schmidt and Brannath, 2014]{SB14}
Schmidt, S. and Brannath, W. (2014).
\newblock Informative simultaneous confidence intervals in hierarchical
  testing.
\newblock {\em Methods of Information in Medicine}, 53:278--283.

\bibitem[Stallard et~al., 2009]{Stallard09}
Stallard, N., Posch, M., Friede, T., Koenig, F., and Brannath, W. (2009).
\newblock Optimal choice of the number of treatments to be included in
  aclinical trial.
\newblock {\em Stat Med}, 28:1321--1338.

\end{thebibliography}

\appendix
\section{Equivalence between formal and intuitive procedure} 

We will show that the confidence bounds $L_{EP}$ and $L_{ER}$ that are defined by the intuitive Figure~\ref{fig_design_intuitive} are identical to the bounds defined by Figure~\ref{fig_design_formal}, i.e., by Formulas~\eqref{lep} and~\eqref{ler}. Note that the quantities $\ell_{EP}$ and $\ell_{ER}$ are equivalent to the test statistics for $H_{EP}^S$ and $H_{ER}^N$, respectively, in the sense that $H_{EP}^S$ is rejected if and only if $\ell_{EP}\ge 0$, and $H_{ER}^N$ is rejected if and only if $\ell_{EP}\ge -\delta_0$. We will discuss all cases.

If $\ell_{EP}<0$, then $H_{EP}^S$ is not rejected and $L_{EP}=\ell_{EP}<0, L_{ER}=-\infty$ with both the formal and the intuitive design, because the first family of hypotheses $(H_{EP}^\vartheta)_{\vartheta\le 0}$ is not rejected completely and both procedures stop. 

If $\ell_{EP}\ge 0$ and $\ell_{ER} < -\delta_0$, then $H_{EP}^S$ is rejected and $H_{ER}^N$ is accepted so that $L_{EP}=0$ and $L_{ER}=\ell_{ER}$ with the formal design. In the intuitive design, we can -- informally written -- see that ``$R > E > P$'', i.e., the filter is satisfied. More formally, we obtain from the definitions of~$\ell_{EP}$ and~$\ell_{ER}$ that
$$
X_R-X_P = \ell_{EP} - (\ell_{ER}+\delta_0) +z_\alpha\sigma\big(\sqrt{n_E^{-1}+n_P^{-1}} - \sqrt{n_E^{-1}+n_R^{-1}}\big) + \delta_0.
$$
Because $\ell_{EP}\ge 0$ and $\ell_{ER}+\delta_0 < 0$, it follows that the filter~\eqref{filterE} holds. Therefore, $L_{ER}=\ell_{ER}$ and $L_{EP}=\max\{L_{ER}+\delta_0,0\}=\max\{\ell_{ER}+\delta_0,0\}=0$, as in the formal design.

If $\ell_{EP}\ge 0$ and $\ell_{ER}\ge -\delta_0$, then we reject both $H_{EP}^S$ and $H_{ER}^N$ with the formal procedure. The confidence bounds are $L_{EP}=\ell_{ER}+\delta_0,L_{ER}=\ell_{ER}$ if the filter~\eqref{filterE} holds and $L_{EP}=\ell_{EP}, L_{ER}=\ell_{EP}-\delta_0$ if the filter does not hold. Consider now the intuitive design. If the filter holds, then $L_{ER}=\ell_{ER}$ and $L_{EP}=\max\{L_{ER}+\delta_0,0\} =\ell_{ER}+\delta_0$. It the filter does not hold, then $L_{EP}=\ell_{EP}$ and $L_{ER}= L_{EP} - \delta_0 = \ell_{EP}-\delta_0$. All bounds correspond to those of the formal design in the respective cases.

\end{document}